\documentclass[sigconf]{acmart}
\usepackage{soul}
\usepackage{siunitx}  % 引入数字格式化宏包

\usepackage{graphicx}
\usepackage{booktabs}
\usepackage{float}
\usepackage{overpic}
\usepackage{algorithm}
\usepackage{algpseudocode}
\usepackage{hyperref}
\usepackage{float}  %设置图片浮动位置的宏包
\usepackage{subfloat}
\usepackage{stfloats} % for pic
\usepackage[skip=0.5\baselineskip]{caption}
\usepackage{bm}
\usepackage{amsmath}
\usepackage{booktabs}
\usepackage{multirow}
\usepackage{multicol}
\usepackage{enumitem}
\usepackage{subcaption}
\usepackage{threeparttable}
\usepackage{xspace}
\usepackage[normalem]{ulem}
\usepackage{algorithmicx,algorithm}
\usepackage{algpseudocode}
\usepackage{algorithm,algpseudocode}
\usepackage{marvosym}   % 添加通讯作者的信封
\usepackage{colortbl}   % 表格中的阴影颜色
\usepackage{tcolorbox}  % 文本加框
\tcbuselibrary{breakable}   % 配合文本加框，使得框可以跨页
\usepackage{wasysym}    % 添加实心圆空心圆
\usepackage[figuresright]{rotating} % 旋转表格
\usepackage{pifont} % 圆圈序号
\newcommand{\ie}{\emph{i.e.,}\xspace}
\newcommand{\eg}{\emph{e.g.,}\xspace}
\newcommand{\name}{ComeIR}
\AtBeginDocument{%
	\providecommand\BibTeX{{%
			\normalfont B\kern-0.5em{\scshape i\kern-0.25em b}\kern-0.8em\TeX}}}

\begin{document}
%\fancyhead{}
\title{Conditional Memory Enhanced Item Representation for Generative Recommendation}
%\author{Anonymous Author(s)}
\author{Ziwei Liu}
% \orcid{0000-0002-0751-2602}
\affiliation{%
  \institution{City University of Hong Kong}
  \city{Hong Kong}
  % \state{Shaanxi}
  \country{China}
}
%\email{lziwei2-c@my.cityu.edu.hk}

\author{Yejing Wang}
% \orcid{0000-0003-2852-9910}
\affiliation{%
  \institution{City University of Hong Kong}
  \city{Hong Kong}
  % \state{Guangdong}
  \country{China}
}
%\email{yejing.wang@my.cityu.edu.hk}

\author{Shengyu Zhou}
\affiliation{%
  \institution{Independent Researcher}
  \city{Beijing}
  % \state{Guangdong}
  \country{China}
}

\author{Xinhang Li}
% \orcid{0000-0003-1417-2295}
\affiliation{%
  \institution{Tsinghua University}
  \city{Beijing}
  % \state{Guangdong}
  \country{China}
}

\author{Xiangyu Zhao\Letter}
% \orcid{0000-0003-2852-9910}
\affiliation{%
  \institution{City University of Hong Kong}
  \city{Hong Kong}
  % \state{Guangdong}
  \country{China}
}

\begin{abstract}
Generative recommendation (GR) has emerged as a promising paradigm that predicts target items by autoregressively generating their semantic identifiers (SID). 
Most GR methods follow a quantization-representation-generation pipeline. While most research focuses on quantization and generation, the representation that bridges the two remains underexplored. 
Existing representation constructions mainly take two forms: either directly flattening the SID code embeddings into a sequence or merging them into a compact vector via external networks. While effective, they still expose two practical challenges: \textbf{1) Loss of item semantics:} Information lost by quantization or compression needs to be recovered during the representation stage. \textbf{2) Insufficient modeling of structured SID relations:} Intra-item code composition patterns and inter-item prefix-transition patterns are not explicitly preserved for representation and subsequent SID generation.

To overcome these challenges, we propose \textbf{\name}, a \textbf{\uline{Co}}nditional \textbf{\uline{me}}mory Enhanced \textbf{\uline{I}}tem \textbf{\uline{R}}epresentation framework that reconstructs SID code embeddings into item-aware inputs and preserves structured SID evidence for prediction. Item-guided Token Scoring estimates the contribution of each SID code using the item embedding employed by the quantizer. Dual-level Conditional Memory captures intra-item prefix composition and inter-item prefix transitions, while Memory Merge integrates these signals into each historical item representation. The Memory-restoring Prediction Head further conditions SID-level prediction on the generated prefix and historical transition evidence. Extensive experiments demonstrate the effectiveness and flexibility of \name{}, and further reveal scalable gains from enlarging conditional memory. %Code is available at {\url{https://anonymous.4open.science/r/ComeIR-WWW27/}}.
\end{abstract}

\setcopyright{acmlicensed}
\copyrightyear{2027}
\acmYear{2027}
\acmDOI{XXXXXXX.XXXXXXX}
%% These commands are for a PROCEEDINGS abstract or paper.
%\acmConference[WWW '27]{Proceedings of the ACM Web Conference 2027 (WWW ’27).}{May 20--14,2027}{Dublin, Ireland}
%%
%%  Uncomment \acmBooktitle if the title of the proceedings is different
%%  from ``Proceedings of ...''!
%%
%%\acmBooktitle{Woodstock '18: ACM Symposium on Neural Gaze Detection,
%%  June 03--05, 2018, Woodstock, NY}
\acmISBN{978-1-4503-XXXX-X/2018/06}

\maketitle
\section{Introduction}\label{sec:intro}
With recent advances in natural language processing, large language models (LLMs) have demonstrated powerful capabilities in both semantic understanding and sequence modeling~\cite{zhao2023survey}, which have motivated a wide range of applications in recommender systems~\cite{zhao2023survey,bai2025bi}. Among the LLM-enhanced solutions, generative recommendation (GR)~\cite{li2024large} has emerged as a prevailing paradigm due to its scalability and outstanding performance in enhancing recommendation diversity~\cite{rajput2023recommender}.

\begin{figure*}[!t]

	\centering

	\includegraphics[width = \linewidth]{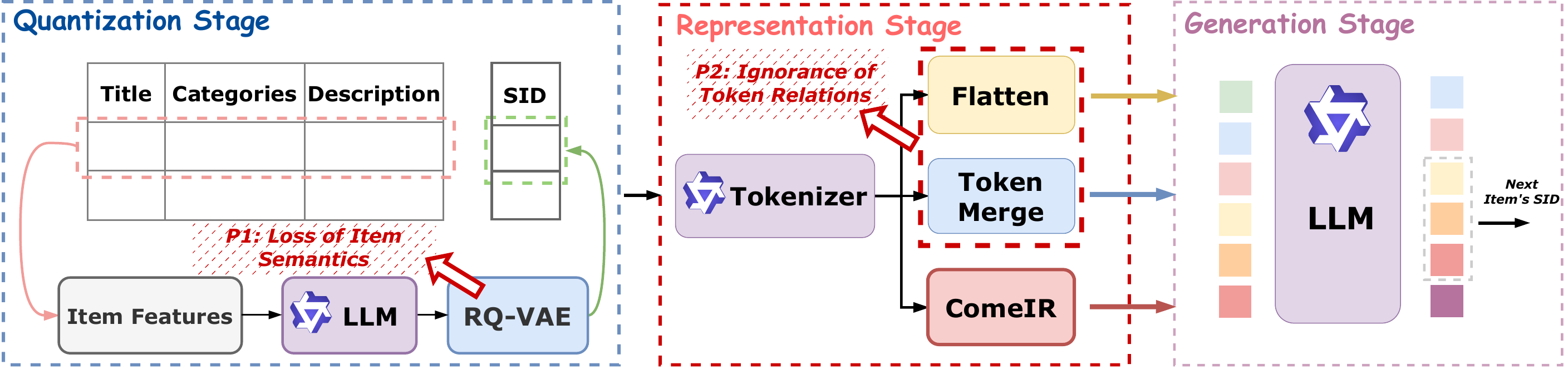}

	\caption{Overview of GR pipeline. Quantization transforms items from features to SID. Representation organizes SID as input to the LLM. Generation uses the LLM to generate the next item's SID conditioned on the organized input. }\label{fig:pre}
\end{figure*}

As shown in the Figure~\ref{fig:pre}, the typical GR paradigm involves three stages~\cite{li2025survey}: (a) quantization~\cite{wang2024learnable,hou2023learning,hou2025generating,hu2026stop,li2026lsig}, (b) Representation~\cite{zhou2025onerec,zou2026genrec,wang2026intrr}, and (c) Generation~\cite{deng2025onerec,lin2025rec,yang2025sparse,mekonnen2026parametric,chen2026beyond}. While quantization and generation proceed by first assigning each item a Semantic ID (SID), \ie, a tuple of discrete codes, and then predicting the next item's SID via autoregressive generation, the representation stage remains underexplored. \emph{Yet this bridge is crucial: since the LLM only observes representations rather than SIDs, this construction determines how much quantized semantics can be preserved for generation}.

Existing representation constructors commonly follow two lines. One directly maps each SID's code to its corresponding token embedding and flattens all such embeddings into a token-level representation~\cite{rajput2023recommender,li2024large,li2025survey}. The other instead constructs the item-level representations by merging the SID code embeddings of each item via an external network~\cite{zhou2025onerec,zhang2026onetrans,zou2026genrec,zhou2025onerec,wang2026intrr}.

Although these strategies appear effective, two challenges still undermine their practical utility: 
\begin{itemize}[leftmargin=*]
    \item \textbf{Loss of item semantics.} In the quantization stage, items are represented as finite discrete codes by training a tokenizer, such as RQ-VAE~\cite{ju2025generative}, which naturally introduces inevitable semantic information loss~\cite{rajput2023recommender}. Existing methods not only fail to recover but amplify this by merging without any guidance, thus deriving coarse representations for GR to consume.

    \item \textbf{Insufficient modeling of structured SID.} SIDs contain two forms of structured evidence: \emph{intra-item code-composition patterns}, which describe how later codes refine an SID prefix, and \emph{inter-item prefix-transition patterns}, which describe recurring transitions among historical item prefixes. Flattening leaves the downstream LLM to infer these SID relations implicitly, whereas other merging methods compress the code-level evidence before generation. Consequently, neither strategy explicitly preserves structured SID relations across representation construction and token-level SID generation.
\end{itemize}

These challenges thus call for a representation construction that is semantics-preserving, 
relation-aware, and compatible with token-level SID generation. To this end, we 
propose a \textbf{\uline{Co}}nditional \textbf{\uline{me}}mory Enhanced 
\textbf{\uline{I}}tem \textbf{\uline{R}}epresentation framework (\textbf{\name}), 
to reconstruct the input of general GR and reuse structured SID evidence during generation. First, \textit{Item-guided Token Scoring} uses the original item embedding consumed by the quantizer to estimate the contribution of each SID code, recovering the lost item semantics during quantization or compression. Second, to preserve the relation of structured SID at both granularities, \ie intra-item and inter-item, \name~optimizes both representation construction and generation. Specifically, it first introduces a \textit{Dual-level Conditional Memory} that models intra-item code composition and inter-item transition patterns. On the representation side, \textit{Memory Merge} adaptively injects the retrieved evidence into each item representation, allowing every historical item to carry both its SID composition and local transition context into the LLM. On the generation side, the \textit{Memory-restoring Prediction Head} reuses the same memory types to provide structured relation evidence during autoregressive decoding. Our contributions can be summarized as follows:
%% token merge 会带来position 缺失的概念，通过score来保留position的实验。 贴attention 分布；原先的token embedding和后续的结合merge之后的embedding； 可视化

\begin{itemize}[leftmargin=*]

    \item We identify representation construction as an underexplored bottleneck in current GR and highlight two limitations of existing representation constructors, \ie Loss of item semantics and Insufficient modeling of structured SID relations.%We identify representation construction as an underexplored bottleneck in current GR, showing that directly flattening or simply compressing SID-token embeddings leads to three severe problems, \ie Item identity ambiguity, SID structure loss, and input-output granularity mismatch.

    \item We propose \textbf{\name}, a plug-and-play conditional memory-enhanced representation framework for generative recommendation that is semantics-preserving, relation-aware, and compatible with token-level SID generation.

    \item %We empirically demonstrate scaling laws in the proposed conditional memory and conduct extensive experiments on three public datasets, validating the effectiveness and generality of \name{}. 
    We conduct experiments on three public datasets to validate the effectiveness and generality of \name{}. Further analysis highlights consistent scalable gains from enlarging the proposed conditional memory for preserving the SID structured relations.

\end{itemize}

\section{Preliminary}\label{sec:preliminary}
The goal of GR is to directly generate the next item that a user is likely to interact with based on historical interactions. Let $\mathcal{U}$ and $\mathcal{I}$ denote the user set and item set, respectively. For each user $u \in \mathcal{U}$, the historical interactions are arranged in chronological order as: $\mathcal{S}_u = \left(v_1, \ldots,v_n,\ldots, v_N\right)$, where $v_n \in \mathcal{I}$ and $N$ denotes the sequence length. For simplicity, we omit the user index $u$ in the following. In the quantization stage, each item is represented by a fixed-length SID rather than a single atomic ID, and the SID of item $v_n$ is
$    \boldsymbol{c}_n
    =
    \left(c_n^1,c_n^2,\ldots,c_n^L\right)$,
where $L$ is the length of the SID. During the representation stage, each SID code is converted into a token embedding by combining its frozen codebook embedding with a learnable token embedding. For the $\ell$-th code of item $v_n$,
\begin{equation}
    \boldsymbol{e}_{c_n^\ell}
    =
    \boldsymbol{W}_{E}
    \left[
    \boldsymbol{e}^{B}_{c_n^\ell};
    \boldsymbol{e}^{L}_{c_n^\ell}
    \right]
    \label{equ:sid_token_embedding}
\end{equation}
where $\boldsymbol{e}^{B}_{c_n^\ell}$ is the frozen codebook embedding, $\boldsymbol{e}^{L}_{c_n^\ell}$ is a learnable token embedding, and $\boldsymbol{W}_{E}$ projects their concatenation into the LLM hidden space. The SID-level representation of item $v_n$ is
$    \boldsymbol{R}_{n}^{S}
    =
    \left[
    \boldsymbol{e}_{c_n^1},
    \ldots,
    \boldsymbol{e}_{c_n^L}
    \right]$.
We can then write the representation construction as
$    \boldsymbol{r}_{n}^{I}
    =
    f
    \left(
    \boldsymbol{R}_{n}^{S},
    \boldsymbol{b}_{n}
    \right)$, and formulate the final GR input as
$    \boldsymbol{R}^{I}
    =
    \left[
    \boldsymbol{r}_{1}^{I},
    \ldots,
    \boldsymbol{r}_{N}^{I}
    \right]$.
Here, $\boldsymbol{b}_{n}$ denotes optional external inputs that provide additional information beyond the SID tokens. When no external inputs are introduced, $\boldsymbol{b}_{n}$ is omitted: for flattening, $f(\cdot)$ reduces to a simple concatenation, while for token merging, $f(\cdot)$ denotes a linear layer. For our \name~, we also omit $\boldsymbol{b}_{n}$ and leverage the existing multi-modal embedding and conditional memories to enhance our representation. Finally, in the generation stage, a generative model $\Theta$ autoregressively predicts the SID of the next item based on the item-level representations, which can be formulated as:
\begin{equation}
    P\left(\boldsymbol{c}_{N+1} \middle| \boldsymbol{R}^{I};\Theta\right)
    =
    \prod_{\ell=1}^{L}
    P\left(c_{N+1}^{\ell} \middle| \boldsymbol{R}^{I}, c_{N+1}^{<\ell};\Theta\right)
\end{equation}

%where $\boldsymbol{e}^{B}_{c_i^\ell}$ is the frozen codebook embedding, $\boldsymbol{e}^{T}_{c_i^\ell}$ is a learnable token embedding, and $\boldsymbol{W}_{E}$ projects their concatenation into the LLM hidden space. We denote the resulting SID-token embedding sequence as $\boldsymbol{E}_{i}^{C}=\left[\boldsymbol{e}_{c_i^1}, \ldots, \boldsymbol{e}_{c_i^L}\right]$. The flattened token embedding sequence can then be written as $\boldsymbol{E}_{I}=\left[\boldsymbol{E}_{1}^{C}, \ldots, \boldsymbol{E}_{N}^{C}\right]$, which is further organized into item-level representations $\boldsymbol{E}^{M}$ to construct the GR input. Finally, in the generation stage, a generative model $\Theta$ autoregressively predicts the SID of the next item based on the constructed representations:

\section{Method}\label{sec:method}
\begin{figure*}[!t]
    \centering
    \includegraphics[width=\linewidth]{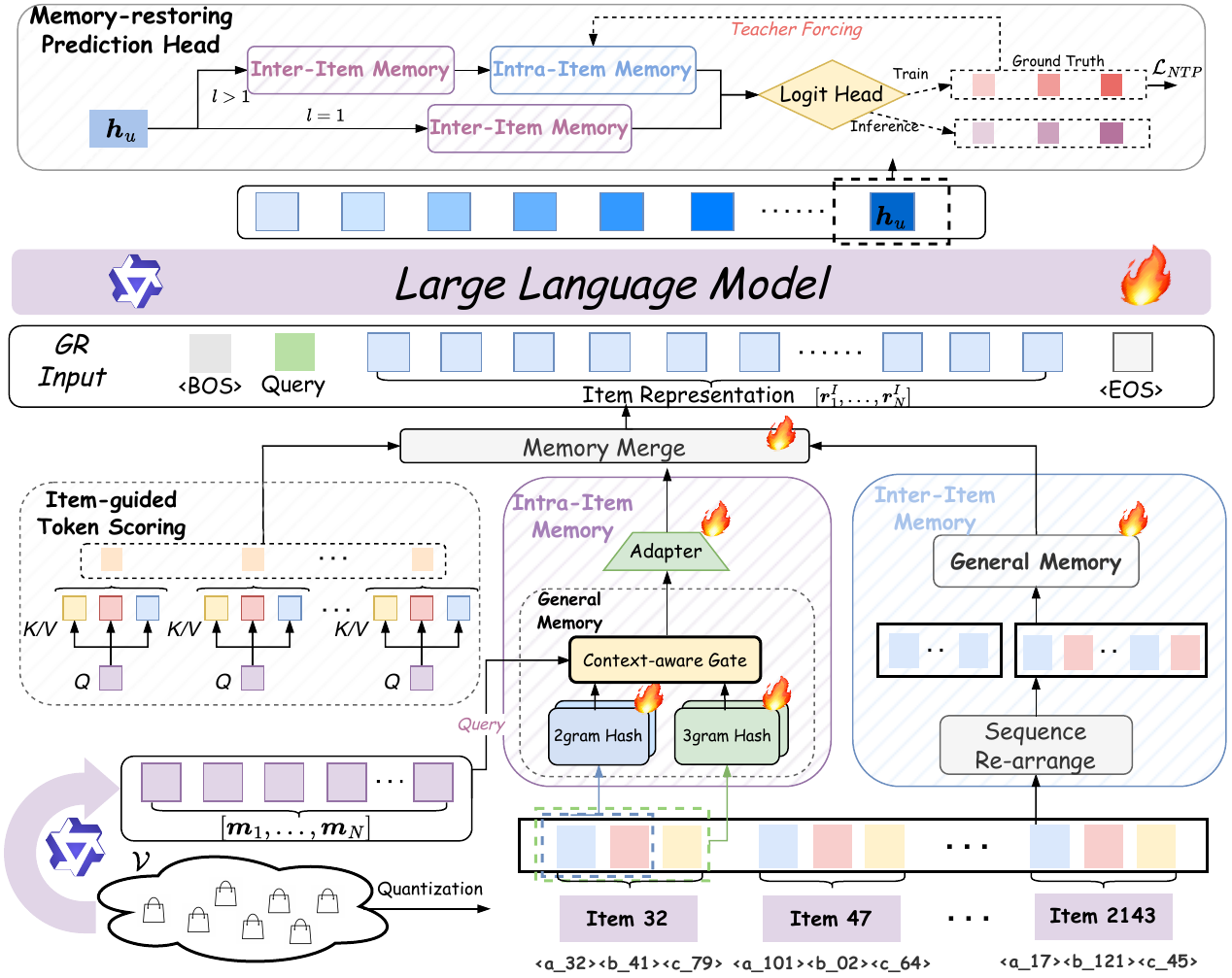}
    \caption{The overall framework of the proposed \name. The SID length is set to $L=3$ for illustration.}\label{fig:overall}
    \vspace{-2mm}
\end{figure*}

\subsection{Framework Overview}
As illustrated in Figure~\ref{fig:overall}, \name~enhances the representation and prediction stages of GR through four components. \textit{Item-guided Token Scoring} estimates the contribution of each SID code and constructs an item-guided semantic context. \textit{Dual-level Conditional Memory} captures intra-item prefix composition and inter-item prefix transitions. \textit{Memory Merge} integrates the resulting level-wise evidence into a compact item representation. Finally, the \textit{Memory-restoring Prediction Head} conditions each SID-level prediction on the generated prefix and historical transition context. The final subsection presents the training and inference procedures.

\subsection{Item-guided Token Scoring}\label{sec:mm_scoring}
Quantization represents an item with multiple discrete codes, whose contributions to the original item semantics may differ across SID levels. To recover this information during representation construction, we propose \textbf{Item-guided Token Scoring}. For item $v_n$, let $\boldsymbol m_n\in\mathbb R^{d_I}$ be the item embedding used to construct its SID. We project it into the LLM hidden space as $\boldsymbol q_n=\boldsymbol W_I\boldsymbol m_n$ and use it to estimate the contribution of each SID-token embedding:
\begin{equation}\label{equ:sid_token_scoring}
\begin{aligned}
\alpha_n^\ell
&=
\frac{
\exp\!\left(\boldsymbol q_n^\top\boldsymbol e_{c_n^\ell}/\sqrt d\right)
}{
\sum_{r=1}^{L}
\exp\!\left(\boldsymbol q_n^\top\boldsymbol e_{c_n^r}/\sqrt d\right)
},\\
\boldsymbol s_n^0
&=\sum_{\ell=1}^{L}\alpha_n^\ell\boldsymbol e_{c_n^\ell}
\in\mathbb R^d.
\end{aligned}
\end{equation}
Here, $\boldsymbol s_n^0$ summarizes the SID tokens according to their relevance to the original item semantics. It subsequently serves as the semantic anchor for conditional-memory retrieval and Memory Merge.

\subsection{Dual-level Conditional Memory}\label{sec:dual_engram}
To model structured SID relations, we consider two discrete pattern types. Within an item, the ordered SID $(c_n^1,\ldots,c_n^L)$ contains intra-item code-composition patterns, where each later code refines its preceding prefix. Across user interactions, the level-$\ell$ prefixes contain inter-item transition patterns that contain collaborative information for characterizing recurring preference changes.

We therefore introduce \textbf{Dual-level Conditional Memory}, consisting of Intra-item Memory and Inter-item Memory. Unlike sparse $n$-gram memory inserted into Transformer blocks~\cite{zhao2017ngram2vec,lin2026collision,cheng2026conditional}, the proposed memories operate both at the representation interface for constructing the GR input and at the generation stage for predicting the next SID. We first define a general conditional-memory read and then instantiate it for the two patterns.

\subsubsection{General Memory.}
Let $X\in\{S,T\}$ denote the memory type, where $S$ represents intra-item SID composition and $T$ represents inter-item transition. A typed memory read receives a discrete pattern $\boldsymbol p$ for addressing and a contextual query $\boldsymbol q$ for determining when the retrieved evidence should be used.

\noindent\emph{\textbf{Step 1: Hashed $n$-gram lookup.}}
For $\boldsymbol p=(p_1,\ldots,p_U)$, an order-$o$ key contains its last $o$ elements, with the element type determined by $X$. Let $\mathcal O_\ell^X$ be the order set at SID level $\ell$. Each key is hashed into $K_X$ sparse tables through multi-head hashing~\citep{tito2017hash}. The vectors retrieved from all orders and heads are concatenated as:
\begin{equation}\label{equ:general_memory_address}
\boldsymbol a_\ell^X(\boldsymbol p)
=
\mathop{\Vert}_{o\in\mathcal O_\ell^X}
\mathop{\Vert}_{k=1}^{K_X}
\widehat{\boldsymbol M}_{\ell,o,k}^{X}
\!\left[\phi_{\ell,o,k}^{X}(g_o(\boldsymbol p))\right],
\end{equation}
where $g_o(\boldsymbol p)$ extracts the order-$o$ suffix and $\widehat{\boldsymbol M}_{\ell,o,k}^{X}$ denotes the corresponding sparse table. Repeated discrete patterns deterministically address the same rows and can accumulate reusable SID evidence.

\noindent\emph{\textbf{Step 2: Context-aware gating.}}
Static $n$-gram patterns may contain context-independent noise caused by hash collisions or pattern ambiguity~\cite{zhao2017ngram2vec,cheng2026conditional}. We use a scalar gate to measure the compatibility between the continuous query and the concatenated sparse address. The typed memory read is
\begin{equation}\label{equ:engram_read}
\begin{aligned}
\boldsymbol v_\ell^X(\boldsymbol p)
&=\boldsymbol W_{V,\ell}^{X}\boldsymbol a_\ell^X(\boldsymbol p),\\
\lambda_\ell^X(\boldsymbol q,\boldsymbol p)
&=\sigma\!\left(
\frac{
(\boldsymbol W_{Q,\ell}^{X}\boldsymbol q)^\top
(\boldsymbol W_{K,\ell}^{X}\boldsymbol a_\ell^X(\boldsymbol p))
}{\sqrt{d_g}}+b_\ell^X
\right),\\
\mathcal R_\ell^X(\boldsymbol q,\boldsymbol p)
&=\operatorname{Norm}^{X}\!\left(
\boldsymbol W_{O,\ell}^{X}
\bigl(\lambda_\ell^X(\boldsymbol q,\boldsymbol p)
\boldsymbol v_\ell^X(\boldsymbol p)\bigr)
\right).
\end{aligned}
\end{equation}
The discrete key determines which recurring SID pattern is retrieved, while the continuous query controls its relevance to the current semantic or transition context. The two memory types use independent tables and order sets; their addressing details are provided in Appendix~\ref{sec:engram_general}.

\subsubsection{Intra-item Memory.}
Intra-item Memory models how deeper SID codes refine their preceding prefixes. For level $t\geq2$, we encode the prefix $\boldsymbol c_n^{\leq t}$ as
\begin{equation}\label{equ:intra_prefix_encoding}
\kappa_{n,t}
=\sum_{j=1}^{t}c_n^j C^{j-1},
\end{equation}
where $C$ is the codebook size. Using the item-guided context as the query, the conditional prefix evidence is
\begin{equation}\label{equ:intra_read}
\boldsymbol z_{n,t}^{S}
=\mathcal R_t^S(\boldsymbol s_n^0,\kappa_{n,t}).
\end{equation}
For representation construction, the reads available up to level $\ell$ are composed into a level-specific intra-item memory:
\begin{equation}\label{equ:intra_memory}
\boldsymbol\eta_{n,\ell}^{S}
=
\operatorname{RMSNorm}\!\left(
\boldsymbol W_{C,\ell}^{S}
[\boldsymbol z_{n,2}^{S};\ldots;\boldsymbol z_{n,\ell}^{S}]
+\boldsymbol b_{C,\ell}^{S}
\right),\quad \ell\geq2,
\end{equation}
with $\boldsymbol\eta_{n,1}^{S}=\boldsymbol0$. In this way, deeper levels accumulate the composition evidence associated with all available SID-prefix refinements.

\subsubsection{Inter-item Memory.}
Inter-item Memory captures prefix-transition patterns across historical items at multiple SID granularities. Shallow prefixes describe coarse preference transitions, whereas deeper prefixes characterize more specific changes. Rather than introducing another prefix encoding, we rearrange the encoded prefixes in Equation~\eqref{equ:intra_prefix_encoding} by SID level and chronological position. At level $\ell$, the resulting sequence is
\begin{equation}\label{equ:inter_prefix_sequence}
\mathcal C^{\leq\ell}
=
(\kappa_{1,\ell},\kappa_{2,\ell},\ldots,\kappa_{N,\ell}),
\qquad
\boldsymbol p_{n,\ell}^{T}
=
\operatorname{Suffix}_{\tau_\ell}
(\mathcal C^{\leq\ell},n),
\end{equation}
where $\tau_\ell\geq\max\mathcal O_\ell^T$. This sequence rearrangement changes the unit read by the general conditional memory from the code prefixes within one item to the level-$\ell$ prefixes across interacted items, while preserving their chronological order. Since a transition also depends on the surrounding interaction context, we construct a level-specific query from a causal local window. Given $\mathcal H_n=\{a\mid\max(1,n-w+1)\leq a\leq n\}$,
\begin{equation}\label{equ:inter_query}
\begin{aligned}
\pi_{n,a,\ell}
&=
\frac{
\exp\!\left(
\frac{
(\boldsymbol W_{Q,\ell}^{P}\boldsymbol s_n^0)^\top
(\boldsymbol W_{K,\ell}^{P}\boldsymbol s_a^0)
}{\sqrt{d_P}}
\right)
}{
\sum_{r\in\mathcal H_n}
\exp\!\left(
\frac{
(\boldsymbol W_{Q,\ell}^{P}\boldsymbol s_n^0)^\top
(\boldsymbol W_{K,\ell}^{P}\boldsymbol s_r^0)
}{\sqrt{d_P}}
\right)
},\\
\boldsymbol q_{n,\ell}^{T}
&=\sum_{a\in\mathcal H_n}\pi_{n,a,\ell}\boldsymbol s_a^0.
\end{aligned}
\end{equation}
The transition memory is then retrieved and locally enhanced as
\begin{equation}\label{equ:inter_memory}
\begin{aligned}
\widetilde{\boldsymbol\eta}_{n,\ell}^{T}
&=\mathcal R_\ell^T(
\boldsymbol q_{n,\ell}^{T},\boldsymbol p_{n,\ell}^{T}),\\
\boldsymbol\eta_{n,\ell}^{T}
&=\widetilde{\boldsymbol\eta}_{n,\ell}^{T}
+\operatorname{ShortConv}_{\ell}
(\widetilde{\boldsymbol\eta}_{1:N,\ell}^{T})_n.
\end{aligned}
\end{equation}
Finally, a learned level attention summarizes the multi-granularity transition evidence as:
\begin{equation}\label{equ:inter_level_aggregation}
\begin{aligned}
\beta_{n,\ell}
&=\operatorname{softmax}_{\ell}
\left(
\boldsymbol w_A^\top
\operatorname{SiLU}(\boldsymbol W_A\boldsymbol\eta_{n,\ell}^{T})
\right),\\
\bar{\boldsymbol\eta}_{n}^{T}
&=\sum_{\ell=1}^{L}\beta_{n,\ell}\boldsymbol\eta_{n,\ell}^{T}.
\end{aligned}
\end{equation}

\subsection{Memory Merge}\label{sec:token_merge}
After obtaining the item-guided context and the two level-wise memories, we need to transform them into one compact item representation. A direct linear merge treats all retrieved patterns equally and may obscure both item semantics and SID structure. We therefore propose \textbf{Memory Merge}, which adaptively injects the memory evidence according to its compatibility with the item-guided semantic anchor derived in Section~\ref{sec:mm_scoring}.

For level $\ell$, we concatenate the two memory signals as $\boldsymbol u_{n,\ell}=[\boldsymbol\eta_{n,\ell}^{S};\boldsymbol\eta_{n,\ell}^{T}]$ and compute
\begin{equation}\label{equ:token_merge}
\begin{aligned}
\omega_{n,\ell}
&=\sigma\!\left(
(\boldsymbol W_Q^{M}\boldsymbol s_n^0)^\top
(\boldsymbol W_{K,\ell}^{M}\boldsymbol u_{n,\ell})
+b_\ell^M
\right),\\
\widetilde{\boldsymbol r}_n^I
&=\operatorname{LN}\!\left(
\boldsymbol s_n^0+
\sum_{\ell=1}^{L}
\omega_{n,\ell}\boldsymbol F_\ell^M(\boldsymbol u_{n,\ell})
\right),\\
\boldsymbol r_n^I
&=\widetilde{\boldsymbol r}_n^I+\bar{\boldsymbol\eta}_n^T,
\end{aligned}
\end{equation}
where $\boldsymbol F_\ell^M$ is a two-layer SiLU MLP. The residual path preserves the item-guided semantics, while the gated updates and aggregated transition residual retain the structured SID evidence. We collect $\boldsymbol R^I=[\boldsymbol r_1^I,\ldots,\boldsymbol r_N^I]$ as the final item-level GR input.

\subsection{Memory-restoring Prediction Head}\label{sec:prediction_head}
While item-level representations reduce the GR input length, the target item must still be predicted as a token-level SID. We therefore introduce the \textbf{Memory-restoring Prediction Head} to recover prefix-composition and historical-transition evidence during level-wise SID prediction.

After encoding the item-level sequence, the final LLM hidden state summarizes the user history:
\begin{equation}\label{equ:vanilla_hidden}
\boldsymbol H^I=\operatorname{LLM}(\boldsymbol R^I),
\qquad
\boldsymbol h_u=\boldsymbol H_N^I.
\end{equation}
At each level, the generated prefix first provides intra-item composition evidence. For every available prefix length $t\in\{2,\ldots,\ell-1\}$, we read $\mathcal R_t^S$ using $\boldsymbol h_u$ and denote the mean of these reads by $\bar{\boldsymbol z}_{u,\ell}^{S}$. The prefix-refined state is
\begin{equation}\label{equ:decode_intra_refine}
\widehat{\boldsymbol h}_{u,\ell}=
\begin{cases}
\operatorname{RMSNorm}(\boldsymbol h_u),&\ell\leq2,\\
\operatorname{RMSNorm}\!\left(
\boldsymbol h_u+
\gamma_{u,\ell}^{S}\boldsymbol W_V^{S,G}
\bar{\boldsymbol z}_{u,\ell}^{S}
\right),&\ell>2,
\end{cases}
\end{equation}
where $\gamma_{u,\ell}^{S}=\sigma((\boldsymbol W_Q^{S,G}\boldsymbol h_u)^\top(\boldsymbol W_K^{S,G}\bar{\boldsymbol z}_{u,\ell}^{S})+b^{S,G})$.

The prediction head then summarizes the historical transition evidence at level $\ell$. For valid historical positions $\mathcal H_u$,
\begin{equation}\label{equ:decode_inter_context}
\begin{aligned}
\rho_{u,a,\ell}
&=
\frac{
\exp\!\left(
(\boldsymbol W_Q^G\boldsymbol h_u)^\top
(\boldsymbol W_K^G\boldsymbol\eta_{a,\ell}^{T})/\sqrt d
\right)
}{
\sum_{r\in\mathcal H_u}
\exp\!\left(
(\boldsymbol W_Q^G\boldsymbol h_u)^\top
(\boldsymbol W_K^G\boldsymbol\eta_{r,\ell}^{T})/\sqrt d
\right)
},\\
\bar{\boldsymbol\eta}_{u,\ell}^{T}
&=\sum_{a\in\mathcal H_u}
\rho_{u,a,\ell}\boldsymbol\eta_{a,\ell}^{T}.
\end{aligned}
\end{equation}
The two restored signals are fused into each prediction state:
\begin{equation}\label{equ:decode_fusion}
\begin{aligned}
\gamma_{u,\ell}^{T}
&=\sigma\!\left(
(\boldsymbol W_Q^{F}\widehat{\boldsymbol h}_{u,\ell})^\top
(\boldsymbol W_K^{F}\bar{\boldsymbol\eta}_{u,\ell}^{T})+b^F
\right),\\
\boldsymbol d_\ell
&=\operatorname{RMSNorm}\!\left(
\widehat{\boldsymbol h}_{u,\ell}+
\gamma_{u,\ell}^{T}\boldsymbol W_V^F
\bar{\boldsymbol\eta}_{u,\ell}^{T}
\right),\\
\boldsymbol\psi_\ell
&=\boldsymbol W_{O,\ell}^{G}\boldsymbol d_\ell.
\end{aligned}
\end{equation}
Here, $\boldsymbol\psi_\ell\in\mathbb R^{|\mathcal C_\ell|}$ contains the logits for all codes at SID level $\ell$.

\subsection{Training and Inference}
\noindent\textbf{Training.}
Given the ground-truth target SID, teacher forcing supplies the preceding codes to the generated-prefix reader. The level-wise probability is
\begin{equation}\label{equ:decode_prob}
P(c_{N+1}^{\ell}=x\mid\boldsymbol R^I,
\boldsymbol c_{N+1}^{<\ell};\Theta)
=
\frac{\exp(\psi_{\ell,x})}
{\sum_{y\in\mathcal C_\ell}\exp(\psi_{\ell,y})},
\end{equation}
and the recommendation objective is
\begin{equation}\label{equ:training_objective}
\mathcal L_{\mathrm{rec}}
=-\sum_{u\in\mathcal U}\sum_{\ell=1}^{L}
\log P(c_{N+1}^{\ell}\mid\boldsymbol R^I,
\boldsymbol c_{N+1}^{<\ell};\Theta).
\end{equation}

\noindent\textbf{Inference.}
During inference, beam search maintains a generated prefix and an accumulated score for each active path. Prefix-conditioned memory reads produce path-specific prediction states, while a catalog prefix trie masks invalid SID extensions after every level. The complete procedure is provided in Appendix~\ref{sec:train_infer}.

\section{Experiment}\label{sec:experiment}
In this section, we conduct extensive experiments on three public datasets to answer the following research questions (RQs).
\begin{itemize}[leftmargin=*]
    \item \textbf{RQ1:} What is the performance of \name~ compared to the SOTA baselines across different quantizers and LLM backbones?

    \item \textbf{RQ2:} How do the proposed modules,\ie Item-guided Token Scoring, Memory Merge, Memory-restoring Prediction Head, and each memory, contribute to \name?

    \item \textbf{RQ3:} What effects does scaling the parameters of the intra-item memory and inter-item memory have?

    \item \textbf{RQ4:} How is the efficiency gains of \name~ compared to normal settings across different LLM backbones? 
    
    \item \textbf{RQ5}: Does \name~preserve hierarchical SID relations for better understanding the historical items of the LLM backbone?
    
\end{itemize}
\subsection{Experimental Settings}\label{sec:settings}

\begin{table*}[t]
\centering
\caption{Overall performance of \name~ under different settings. \textbf{bold} values are the best, and ``*'' marks significant gains (one-side t-test with p<0.05) over the runner-up architecture.}
\label{Tab: Overall_Performance}
\small
\begin{subtable}[t]{\linewidth}
\centering
\caption{Performance with \textbf{Qwen3-0.6B} as the LLM backbone.}
\label{Tab: Overall_Performance_Qwen}
\resizebox{0.99\linewidth}{!}{
\begin{tabular}{cc|cccc|cccc|cccc}
\toprule
\multirow{2}{*}{\textbf{Quantization}}
& \multirow{2}{*}{\textbf{Method}}
& \multicolumn{4}{c|}{\textbf{Yelp}}
& \multicolumn{4}{c|}{\textbf{Industrial}}
& \multicolumn{4}{c}{\textbf{Instrument}} \\
\cmidrule(lr){3-6} \cmidrule(lr){7-10} \cmidrule(lr){11-14}
&
& {H@5} & {H@10} & {N@5} & {N@10}
& {H@5} & {H@10} & {N@5} & {N@10}
& {H@5} & {H@10} & {N@5} & {N@10} \\
\midrule

\multirow{8}{*}{\textbf{RQ-VAE}}
& {OneRec-v2} & 0.0298 & 0.0481 & 0.0185 & 0.0249 & 0.0984 & 0.1284 & 0.0751 & 0.0853 & 0.0810 & 0.0985 & 0.0691 & 0.0740 \\
& {IntRR} & 0.0299 & 0.0488 & 0.0190 & 0.0253 & 0.0991 & 0.1292 & 0.0765 & 0.0864 & 0.0813 & 0.0989 & 0.0693 & 0.0745 \\
& {Vanilla GR} & 0.0296 & 0.0474 & 0.0181 & 0.0243 & 0.0979 & 0.1275 & 0.0745 & 0.0844 & 0.0804 & 0.0982 & 0.0688 & 0.0735 \\
& \textit{+ TM} & 0.0292 & 0.0470 & 0.0177 & 0.0239 & 0.0976 & 0.1271 & 0.0742 & 0.0841 & 0.0788 & 0.0966 & 0.0672 & 0.0719 \\
& \cellcolor{cyan!20} \textit{\textbf{+ \name}} & \cellcolor{cyan!20}\textbf{0.0305\(^*\)} & \cellcolor{cyan!20}\textbf{0.0524\(^*\)} & \cellcolor{cyan!20}\textbf{0.0198\(^*\)} & \cellcolor{cyan!20}\textbf{0.0266\(^*\)} & \cellcolor{cyan!20}\textbf{0.1031\(^*\)} & \cellcolor{cyan!20}\textbf{0.1321\(^*\)} & \cellcolor{cyan!20}\textbf{0.0792\(^*\)} & \cellcolor{cyan!20}\textbf{0.0881\(^*\)} & \cellcolor{cyan!20}\textbf{0.0834\(^*\)} & \cellcolor{cyan!20}\textbf{0.1021\(^*\)} & \cellcolor{cyan!20}\textbf{0.0709\(^*\)} & \cellcolor{cyan!20}\textbf{0.0772\(^*\)} \\
& {NEZHA} & 0.0294 & 0.0473 & 0.0176 & 0.0241 & 0.0981 & 0.1278 & 0.0749 & 0.0847 & 0.0801 & 0.0977 & 0.0684 & 0.0730 \\
& \textit{+ TM} & 0.0287 & 0.0466 & 0.0169 & 0.0234 & 0.0977 & 0.1274 & 0.0746 & 0.0844 & 0.0784 & 0.0960 & 0.0667 & 0.0713 \\
& \cellcolor{cyan!20} \textit{\textbf{+ \name}} & \cellcolor{cyan!20}\textbf{0.0302\(^*\)} & \cellcolor{cyan!20}\textbf{0.0520\(^*\)} & \cellcolor{cyan!20}\textbf{0.0194\(^*\)} & \cellcolor{cyan!20}\textbf{0.0263\(^*\)} & \cellcolor{cyan!20}\textbf{0.1035\(^*\)} & \cellcolor{cyan!20}\textbf{0.1324\(^*\)} & \cellcolor{cyan!20}\textbf{0.0796\(^*\)} & \cellcolor{cyan!20}\textbf{0.0883\(^*\)} & \cellcolor{cyan!20}\textbf{0.0831\(^*\)} & \cellcolor{cyan!20}\textbf{0.1015\(^*\)} & \cellcolor{cyan!20}\textbf{0.0705\(^*\)} & \cellcolor{cyan!20}\textbf{0.0767\(^*\)} \\

\cmidrule(lr){1-14}

\multirow{8}{*}{\textbf{RQ-KMeans}}
& {OneRec-v2}\( {}\) & 0.0297 & 0.0486 & 0.0196 & 0.0257 & 0.1011 & 0.1307 & 0.0777 & 0.0879 & 0.0823 & 0.1003 & 0.0709 & 0.0761 \\
& {IntRR}\( {}\) & 0.0298 & 0.0493 & 0.0201 & 0.0261 & 0.1018 & 0.1315 & 0.0791 & 0.0890 & 0.0826 & 0.1007 & 0.0711 & 0.0766 \\
& {Vanilla GR} & 0.0300 & 0.0481 & 0.0192 & 0.0251 & 0.1005 & 0.1299 & 0.0771 & 0.0870 & 0.0817 & 0.0999 & 0.0705 & 0.0754 \\
& \textit{+ TM} & 0.0295 & 0.0476 & 0.0187 & 0.0246 & 0.1001 & 0.1295 & 0.0767 & 0.0867 & 0.0801 & 0.0983 & 0.0689 & 0.0738 \\
& \cellcolor{cyan!20} \textit{\textbf{+ \name}} & \cellcolor{cyan!20}\textbf{0.0309\(^*\)} & \cellcolor{cyan!20}\textbf{0.0530\(^*\)} & \cellcolor{cyan!20}\textbf{0.0210\(^*\)} & \cellcolor{cyan!20}\textbf{0.0274\(^*\)} & \cellcolor{cyan!20}\textbf{0.1059\(^*\)} & \cellcolor{cyan!20}\textbf{0.1345\(^*\)} & \cellcolor{cyan!20}\textbf{0.0819\(^*\)} & \cellcolor{cyan!20}\textbf{0.0908\(^*\)} & \cellcolor{cyan!20}\textbf{0.0848\(^*\)} & \cellcolor{cyan!20}\textbf{0.1039\(^*\)} & \cellcolor{cyan!20}\textbf{0.0727\(^*\)} & \cellcolor{cyan!20}\textbf{0.0792\(^*\)} \\
& {NEZHA} & 0.0288 & 0.0476 & 0.0187 & 0.0248 & 0.1008 & 0.1299 & 0.0775 & 0.0872 & 0.0814 & 0.0996 & 0.0702 & 0.0753 \\
& \textit{+ TM} & 0.0282 & 0.0470 & 0.0181 & 0.0242 & 0.1004 & 0.1295 & 0.0771 & 0.0869 & 0.0797 & 0.0979 & 0.0685 & 0.0736 \\
& \cellcolor{cyan!20} \textit{\textbf{+ \name}} & \cellcolor{cyan!20}\textbf{0.0297\(^*\)} & \cellcolor{cyan!20}\textbf{0.0524\(^*\)} & \cellcolor{cyan!20}\textbf{0.0204\(^*\)} & \cellcolor{cyan!20}\textbf{0.0271\(^*\)} & \cellcolor{cyan!20}\textbf{0.1062\(^*\)} & \cellcolor{cyan!20}\textbf{0.1345\(^*\)} & \cellcolor{cyan!20}\textbf{0.0823\(^*\)} & \cellcolor{cyan!20}\textbf{0.0909\(^*\)} & \cellcolor{cyan!20}\textbf{0.0845\(^*\)} & \cellcolor{cyan!20}\textbf{0.1035\(^*\)} & \cellcolor{cyan!20}\textbf{0.0723\(^*\)} & \cellcolor{cyan!20}\textbf{0.0791\(^*\)} \\

\bottomrule
\end{tabular}
}
\end{subtable}

\begin{subtable}[t]{\linewidth}
\centering
\caption{Performance with \textbf{LLaMA3-1B} as the LLM backbone.}
\label{Tab: Overall_Performance_LLaMA}
\resizebox{0.99\linewidth}{!}{
\begin{tabular}{cc|cccc|cccc|cccc}
\toprule
\multirow{2}{*}{\textbf{Quantization}}
& \multirow{2}{*}{\textbf{Method}}
& \multicolumn{4}{c|}{\textbf{Yelp}}
& \multicolumn{4}{c|}{\textbf{Industrial}}
& \multicolumn{4}{c}{\textbf{Instrument}} \\
\cmidrule(lr){3-6} \cmidrule(lr){7-10} \cmidrule(lr){11-14}
&
& {H@5} & {H@10} & {N@5} & {N@10}
& {H@5} & {H@10} & {N@5} & {N@10}
& {H@5} & {H@10} & {N@5} & {N@10} \\
\midrule

\multirow{8}{*}{\textbf{RQ-VAE}}
& {OneRec-v2}\( {}\) & 0.0217 & 0.0380 & 0.0135 & 0.0187 & 0.0943 & 0.1231 & 0.0706 & 0.0804 & 0.0755 & 0.0936 & 0.0650 & 0.0702 \\
& {IntRR}\( {}\) & 0.0218 & 0.0387 & 0.0140 & 0.0191 & 0.0950 & 0.1239 & 0.0720 & 0.0815 & 0.0758 & 0.0940 & 0.0652 & 0.0707 \\
& {Vanilla GR} & 0.0216 & 0.0373 & 0.0131 & 0.0182 & 0.0937 & 0.1223 & 0.0701 & 0.0794 & 0.0748 & 0.0932 & 0.0646 & 0.0696 \\
& \textit{+ TM} & 0.0205 & 0.0362 & 0.0120 & 0.0171 & 0.0932 & 0.1218 & 0.0696 & 0.0788 & 0.0731 & 0.0915 & 0.0629 & 0.0679 \\
& \cellcolor{cyan!20} \textit{\textbf{+ \name}} & \cellcolor{cyan!20}\textbf{0.0219\(^*\)} & \cellcolor{cyan!20}\textbf{0.0416\(^*\)} & \cellcolor{cyan!20}\textbf{0.0143\(^*\)} & \cellcolor{cyan!20}\textbf{0.0199\(^*\)} & \cellcolor{cyan!20}\textbf{0.0987\(^*\)} & \cellcolor{cyan!20}\textbf{0.1267\(^*\)} & \cellcolor{cyan!20}\textbf{0.0746\(^*\)} & \cellcolor{cyan!20}\textbf{0.0829\(^*\)} & \cellcolor{cyan!20}\textbf{0.0777\(^*\)} & \cellcolor{cyan!20}\textbf{0.0970\(^*\)} & \cellcolor{cyan!20}\textbf{0.0666\(^*\)} & \cellcolor{cyan!20}\textbf{0.0732\(^*\)} \\
& {NEZHA} & 0.0212 & 0.0372 & 0.0126 & 0.0177 & 0.0941 & 0.1223 & 0.0703 & 0.0798 & 0.0747 & 0.0929 & 0.0644 & 0.0693 \\
& \textit{+ TM} & 0.0200 & 0.0360 & 0.0114 & 0.0165 & 0.0935 & 0.1217 & 0.0697 & 0.0792 & 0.0729 & 0.0911 & 0.0626 & 0.0675 \\
& \cellcolor{cyan!20} \textit{\textbf{+ \name}} & \cellcolor{cyan!20}\textbf{0.0215\(^*\)} & \cellcolor{cyan!20}\textbf{0.0414\(^*\)} & \cellcolor{cyan!20}\textbf{0.0138\(^*\)} & \cellcolor{cyan!20}\textbf{0.0194\(^*\)} & \cellcolor{cyan!20}\textbf{0.0992\(^*\)} & \cellcolor{cyan!20}\textbf{0.1267\(^*\)} & \cellcolor{cyan!20}\textbf{0.0748\(^*\)} & \cellcolor{cyan!20}\textbf{0.0833\(^*\)} & \cellcolor{cyan!20}\textbf{0.0776\(^*\)} & \cellcolor{cyan!20}\textbf{0.0966\(^*\)} & \cellcolor{cyan!20}\textbf{0.0664\(^*\)} & \cellcolor{cyan!20}\textbf{0.0728\(^*\)} \\

\cmidrule(lr){1-14}

\multirow{8}{*}{\textbf{RQ-KMeans}}
& {OneRec-v2}\( {}\) & 0.0225 & 0.0398 & 0.0156 & 0.0213 & 0.0966 & 0.1253 & 0.0729 & 0.0831 & 0.0763 & 0.0952 & 0.0666 & 0.0721 \\
& {IntRR}\( {}\) & 0.0226 & 0.0405 & 0.0161 & 0.0217 & 0.0973 & 0.1261 & 0.0743 & 0.0842 & 0.0766 & 0.0956 & 0.0668 & 0.0726 \\
& {Vanilla GR} & 0.0225 & 0.0393 & 0.0150 & 0.0209 & 0.0961 & 0.1243 & 0.0722 & 0.0821 & 0.0757 & 0.0949 & 0.0662 & 0.0716 \\
& \textit{+ TM} & 0.0216 & 0.0384 & 0.0141 & 0.0200 & 0.0958 & 0.1240 & 0.0719 & 0.0818 & 0.0739 & 0.0931 & 0.0644 & 0.0698 \\
& \cellcolor{cyan!20} \textit{\textbf{+ \name}} & \cellcolor{cyan!20}\textbf{0.0230\(^*\)} & \cellcolor{cyan!20}\textbf{0.0438\(^*\)} & \cellcolor{cyan!20}\textbf{0.0164\(^*\)} & \cellcolor{cyan!20}\textbf{0.0228\(^*\)} & \cellcolor{cyan!20}\textbf{0.1013\(^*\)} & \cellcolor{cyan!20}\textbf{0.1289\(^*\)} & \cellcolor{cyan!20}\textbf{0.0769\(^*\)} & \cellcolor{cyan!20}\textbf{0.0858\(^*\)} & \cellcolor{cyan!20}\textbf{0.0786\(^*\)} & \cellcolor{cyan!20}\textbf{0.0987\(^*\)} & \cellcolor{cyan!20}\textbf{0.0681\(^*\)} & \cellcolor{cyan!20}\textbf{0.0752\(^*\)} \\
& {NEZHA} & 0.0218 & 0.0387 & 0.0149 & 0.0202 & 0.0962 & 0.1248 & 0.0727 & 0.0826 & 0.0754 & 0.0944 & 0.0660 & 0.0711 \\
& \textit{+ TM} & 0.0208 & 0.0377 & 0.0139 & 0.0192 & 0.0959 & 0.1245 & 0.0724 & 0.0823 & 0.0736 & 0.0926 & 0.0642 & 0.0693 \\
& \cellcolor{cyan!20} \textit{\textbf{+ \name}} & \cellcolor{cyan!20}\textbf{0.0223\(^*\)} & \cellcolor{cyan!20}\textbf{0.0431\(^*\)} & \cellcolor{cyan!20}\textbf{0.0163\(^*\)} & \cellcolor{cyan!20}\textbf{0.0221\(^*\)} & \cellcolor{cyan!20}\textbf{0.1015\(^*\)} & \cellcolor{cyan!20}\textbf{0.1294\(^*\)} & \cellcolor{cyan!20}\textbf{0.0774\(^*\)} & \cellcolor{cyan!20}\textbf{0.0863\(^*\)} & \cellcolor{cyan!20}\textbf{0.0784\(^*\)} & \cellcolor{cyan!20}\textbf{0.0982\(^*\)} & \cellcolor{cyan!20}\textbf{0.0680\(^*\)} & \cellcolor{cyan!20}\textbf{0.0747\(^*\)} \\

\bottomrule
\end{tabular}
}
\end{subtable}
\vspace{-3mm}
\end{table*}

\subsubsection{Dataset} We adopted three commonly used datasets for evaluation, \ie Yelp\footnote{\url{https://www.yelp.com/dataset}}, Amazon Industrial\footnote{\url{https://cseweb.ucsd.edu/~jmcauley/datasets.html\#amazon_reviews}}, and Amazon Instrument\footnote{\url{https://cseweb.ucsd.edu/~jmcauley/datasets.html\#amazon_reviews}}. More details can be found in Appendix~\ref{sec:data}.

\subsubsection{Baselines \& Backbones} We construct our experiments compared to powerful item representation constructor, \ie OneRec-v2~\cite{zhou2025onerec} and IntRR~\cite{wang2026intrr} based on the vanilla GR, leveraging different quantization mechanisms, \ie RQ-VAE~\cite{lee2022autoregressive} and RQ-Kmeans~\cite{jegou2010product}, different LLM backbones, \ie Qwen3-0.6B and LLaMA3-1B, and different architectures, \ie Vanilla GR~\cite{rajput2023recommender} and NEZHA~\cite{wang2026nezha}. The `\textit{+ TM}' denotes token merging, following the settings of previous work~\cite{zou2026genrec}, which naively compresses all code embeddings into one compact embedding for each item via a linear layer.

\subsubsection{Evaluation Metrics}
We adopt the commonly used metrics, \ie \textit{hit rate} ($\mathrm{H@K}$) and \textit{Normalized Discounted Cumulative Gain} ($\mathrm{N@K}$), truncated at $K$, where $K \in {5,10}$. For efficiency, we also report the generation latency
(LT) in milliseconds (ms) per batch, and Peak GPU memory in gigabytes (GB) during inference~\cite{wang2026nezha}. 

\subsubsection{Implementation Details}
Experiments run on an AMD EPYC 9745 platform with two NVIDIA RTX PRO 6000 GPUs (Blackwell, 96 GB), Python 3.11, and PyTorch 2.10. The cached $\boldsymbol m_n$ is the embedding supplied to the quantizer when constructing the SID. For general settings, \eg the intra-item and inter-item sparse memory tables, we set a base of 128 and 16, respectively. The two scaling parameters (intra \& inter) for enlarging the table scale are 1.0 and 2.0 by default. Other details can be found in the {Appendix~\ref {sec:implementation_details}}. 

\subsection{Overall Performance (RQ1)}
%To validate the effectiveness and flexibility of the proposed \name, we compare its performance across various quantization mechanisms, LLM backbones, and architectures, both with the general setting (flattening the sequence) and with token merging. As shown in Table~\ref{Tab: Overall_Performance}, \name~demonstrates superior recommendation performance and remarkable robustness. By integrating \name~ into current architectures (\textbf{+ \name}), significant gains are consistently achieved across all commonly adopted quantization mechanisms, \ie RQ-VAE and RQ-KMeans; both backbones, \ie normal GR and NEZHA; and other item-level representation constructors, \ie OneRec-v2 and IntRR. Moreover, the comparison with \textit{+ TM} further shows that simple compression is insufficient; memory-conditioned construction is necessary to preserve fine-grained code evidence during item-level compression. Specifically, on the Yelp dataset with RQ-VAE, \name~yields its largest average improvements over different architectures: \textbf{8.06\%} for Qwen3-0.6B and \textbf{7.91\%} for LLaMA3-1B. The gains are more moderate in Industrial and Instrument, as expected under the fixed-scale setting in Section~\ref{sec:settings}: larger datasets reduce the effective memory capacity, thereby increasing hash collisions and limiting the marginal gain. Overall, the consistent performance gains of \name~ are remarkable, making it a plug-and-play framework that can be seamlessly integrated into various GR pipelines.

%\yy{
To validate the effectiveness and flexibility of the proposed \name, we compare its performance across various quantization mechanisms, LLM backbones, and architectures, both with the general setting (flattening the sequence) and with token merging. Table~\ref{Tab: Overall_Performance} shows that:
\textbf{1) \name~achieves superior recommendation performance with strong robustness.} The \textbf{+ \name} variants consistently yield significant gains across RQ-VAE and RQ-KMeans quantization and both Vanilla GR and NEZHA architectures, while also outperforming the standalone OneRec-v2 and IntRR constructors.
\textbf{2) Comparing + \name~with \textit{+ TM} shows that simple compression is insufficient.} Preserving fine-grained code evidence during item-level compression requires memory-conditioned construction. On Yelp with RQ-VAE, \name~achieves its largest average gains across architectures: \textbf{8.06\%} with Qwen3-0.6B and \textbf{7.91\%} with LLaMA3-1B. Gains on Industrial and Instrument are more moderate under the fixed-scale setting in Section~\ref{sec:settings}, as larger datasets reduce effective memory capacity, increasing hash collisions and limiting marginal gains. These consistent improvements support \name~as a plug-and-play framework for diverse GR pipelines.
%}

\begin{table}[t]
    \centering
    \caption{Ablation results on Yelp. \textit{+ TM} is the no-scoring/no-memory reference. \textit{w/o} Token Scoring uses the same linear constructor as \textit{+ TM} while retaining all memories; \textit{w/o} Mem. Merge replaces the gated merge with a linear layer. \textit{w/o} \textbf{R}-Memory and \textit{w/o} \textbf{G}-Memory remove memories from representation and generation, respectively. Other \textbf{R}/\textbf{G} variants remove one memory type, and Matched Dense Head replaces the memory-restoring head with a size-matched FFN.}
    \label{Tab: Ablation_Main}
    \small
    \resizebox{\columnwidth}{!}{
    \begin{tabular}{ccccc}
    \toprule
    \textbf{Variant} & \textbf{H@5} & \textbf{H@10} & \textbf{N@5} & \textbf{N@10} \\
    \midrule
    \cellcolor{cyan!20}\textbf{\name} & \cellcolor{cyan!20}\textbf{0.0305} & \cellcolor{cyan!20}\textbf{0.0524} & \cellcolor{cyan!20}\textbf{0.0198} & \cellcolor{cyan!20}\textbf{0.0266} \\
    \textit{+ TM} & 0.0292 & 0.0470 & 0.0177 & 0.0239 \\
    \midrule
    \textit{w/o} Token Scoring & 0.0296 & 0.0507 & 0.0191 & 0.0255 \\
    \textit{w/o} Memory Merge & 0.0298 & 0.0502 & 0.0192 & 0.0254 \\
    \textit{w/o} \textbf{R}-intra & 0.0300 & 0.0510 & 0.0193 & 0.0258 \\
    \textit{w/o} \textbf{R}-inter & 0.0302 & 0.0515 & 0.0196 & 0.0261 \\
    \textit{w/o} \textbf{R}-Memory & 0.0296 & 0.0498 & 0.0189 & 0.0251 \\
    \textit{w/o} \textbf{G}-intra & 0.0297 & 0.0504 & 0.0191 & 0.0255 \\
    \textit{w/o} \textbf{G}-inter & 0.0300 & 0.0511 & 0.0195 & 0.0260 \\
    \textit{w/o} \textbf{G}-Memory & 0.0294 & 0.0492 & 0.0186 & 0.0248 \\
    \textit{w/o} All Memory & 0.0297 & 0.0486 & 0.0186 & 0.0248 \\
    Matched Dense Head & 0.0299 & 0.0506 & 0.0192 & 0.0256 \\
    \bottomrule
    \end{tabular}
    }
    \end{table}
\subsection{Ablation Study (RQ2)}
The results are shown in Table~\ref{Tab: Ablation_Main}. We can observe that: \textbf{1): Item guidance and conditional memory are complementary.} Relative to \textit{+ TM}, retaining only Item-guided Token Scoring (\textit{w/o} All Memory) improves all four metrics, while retaining memory with the TM constructor (\textit{w/o} Token Scoring) provides a larger gain. Combining both produces the full \name, showing that the improvement cannot be attributed to either item guidance or memory alone. The degradation from replacing the gated Memory Merge further shows that adaptive memory injection is preferable to unconditional linear aggregation. \textbf{2): Memory is useful on both sides of the GR pipeline.} Removing both representation-side memories (\textit{w/o} \textbf{R}-Memory) or both generation-side memories (\textit{w/o} \textbf{G}-Memory) consistently degrades performance, validating preservation before the LLM and restoration during SID decoding, respectively. \textbf{3): Intra-item composition is the stronger signal, while inter-item transitions remain complementary.} On both the representation and generation sides, removing intra-item memory causes a larger average decline than removing inter-item memory. This is consistent with intra-item memory directly modeling the hierarchical composition required to reconstruct the target SID, whereas inter-item memory supplements transition evidence that can be partially modeled by the LLM. Finally, the size-matched dense alternative remains inferior to the memory-restoring head, supporting the utility of conditional retrieval beyond merely increasing prediction-head capacity. For more detailed results on other datasets, please refer to Appendix~\ref{sec:ablation_supp}.
\begin{table}[t]
\centering
\vspace{-4mm}
\caption{Inference latency (LT) of \name~per batch across different datasets, LLM backbones, and architectures. All baselines use a batch size of 32 and a beam size of 20.}
\label{Tab: LLM_Inference_LT}
\small
\setlength{\tabcolsep}{4pt}
\renewcommand{\arraystretch}{1.15}
\resizebox{\linewidth}{!}{
\begin{tabular}{cc|cc|cc}
\toprule
\multirow{3}{*}{\textbf{Dataset}}
& \multirow{3}{*}{\textbf{Method}}
& \multicolumn{2}{c|}{\textbf{Qwen3-0.6B}}
& \multicolumn{2}{c}{\textbf{LLaMA3-1B}} \\
\cmidrule(lr){3-4} \cmidrule(lr){5-6}
& & {Vanilla GR} & {NEZHA}
& {Vanilla GR} & {NEZHA} \\
& & {LT} $\downarrow$ & {LT} $\downarrow$
& {LT} $\downarrow$ & {LT} $\downarrow$ \\
\midrule

\multirow{4}{*}{\textbf{Yelp}}
& Flattened
& 23.673 & 1.126
& 35.483 & 1.704 \\
& \textit{+ TM}
& \textbf{7.684} & \textbf{0.398}
& \textbf{12.641} & \textbf{0.645} \\
& IntRR
& 11.734 & 0.509
& 16.608 & 0.757 \\
& \cellcolor{cyan!20}\textit{\textbf{ \name}}
& \cellcolor{cyan!20}\underline{9.161}
& \cellcolor{cyan!20}\underline{0.438}
& \cellcolor{cyan!20}\underline{14.322}
& \cellcolor{cyan!20}\underline{0.682} \\
\midrule

\multirow{4}{*}{\textbf{Industrial}}
& Flattened
& 21.406 & 1.018
& 32.109 & 1.538 \\
& \textit{+ TM}
& \textbf{7.326} & \textbf{0.397}
& \textbf{12.884} & \textbf{0.628} \\
& IntRR
& 11.396 & 0.501
& 16.921 & 0.737 \\
& \cellcolor{cyan!20}\textit{\textbf{ \name}}
& \cellcolor{cyan!20}\underline{9.081}
& \cellcolor{cyan!20}\underline{0.432}
& \cellcolor{cyan!20}\underline{14.176}
& \cellcolor{cyan!20}\underline{0.674} \\
\midrule

\multirow{4}{*}{\textbf{Instrument}}
& Flattened
& 26.375 & 1.253
& 39.563 & 1.894 \\
& \textit{+ TM}
& \textbf{7.703} & \textbf{0.395}
& \textbf{13.006} & \textbf{0.641} \\
& IntRR
& 11.628 & 0.520
& 17.268 & 0.747 \\
& \cellcolor{cyan!20}\textit{\textbf{ \name}}
& \cellcolor{cyan!20}\underline{8.874}
& \cellcolor{cyan!20}\underline{0.443}
& \cellcolor{cyan!20}\underline{14.845}
& \cellcolor{cyan!20}\underline{0.680} \\
\bottomrule
\end{tabular}
}
\vspace{-3mm}
\end{table}

\begin{figure}[t]
    \centering
    \captionsetup[subfigure]{font=small,labelfont=bf,justification=centering}
    \begin{subfigure}[t]{\linewidth}
        \centering
        \includegraphics[width=\linewidth]{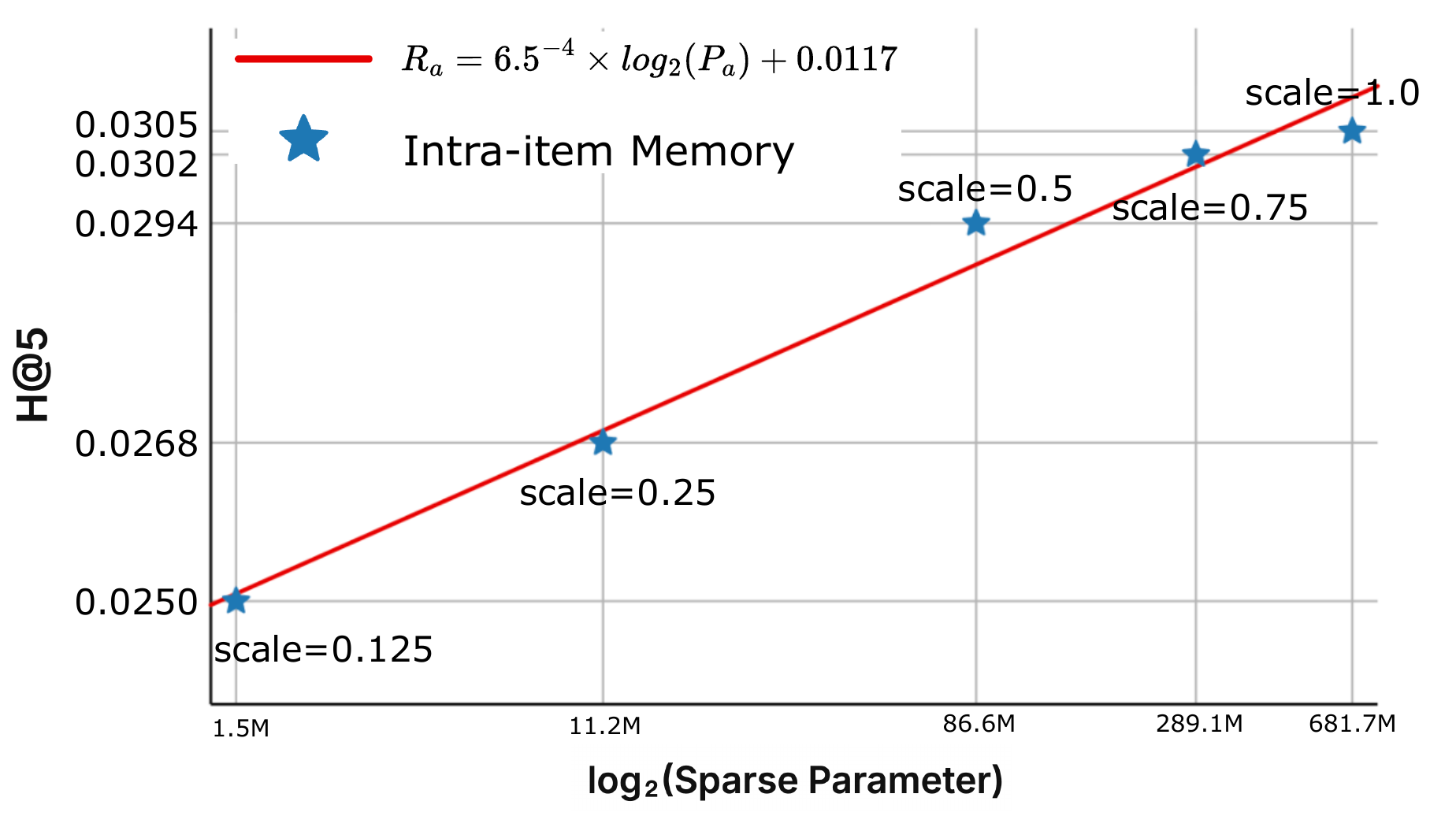}
        \caption{Intra-item Memory}
        \label{fig:intra_scaling}
    \end{subfigure}
    \hfill
    \begin{subfigure}[t]{\linewidth}
        \centering
        \includegraphics[width=\linewidth]{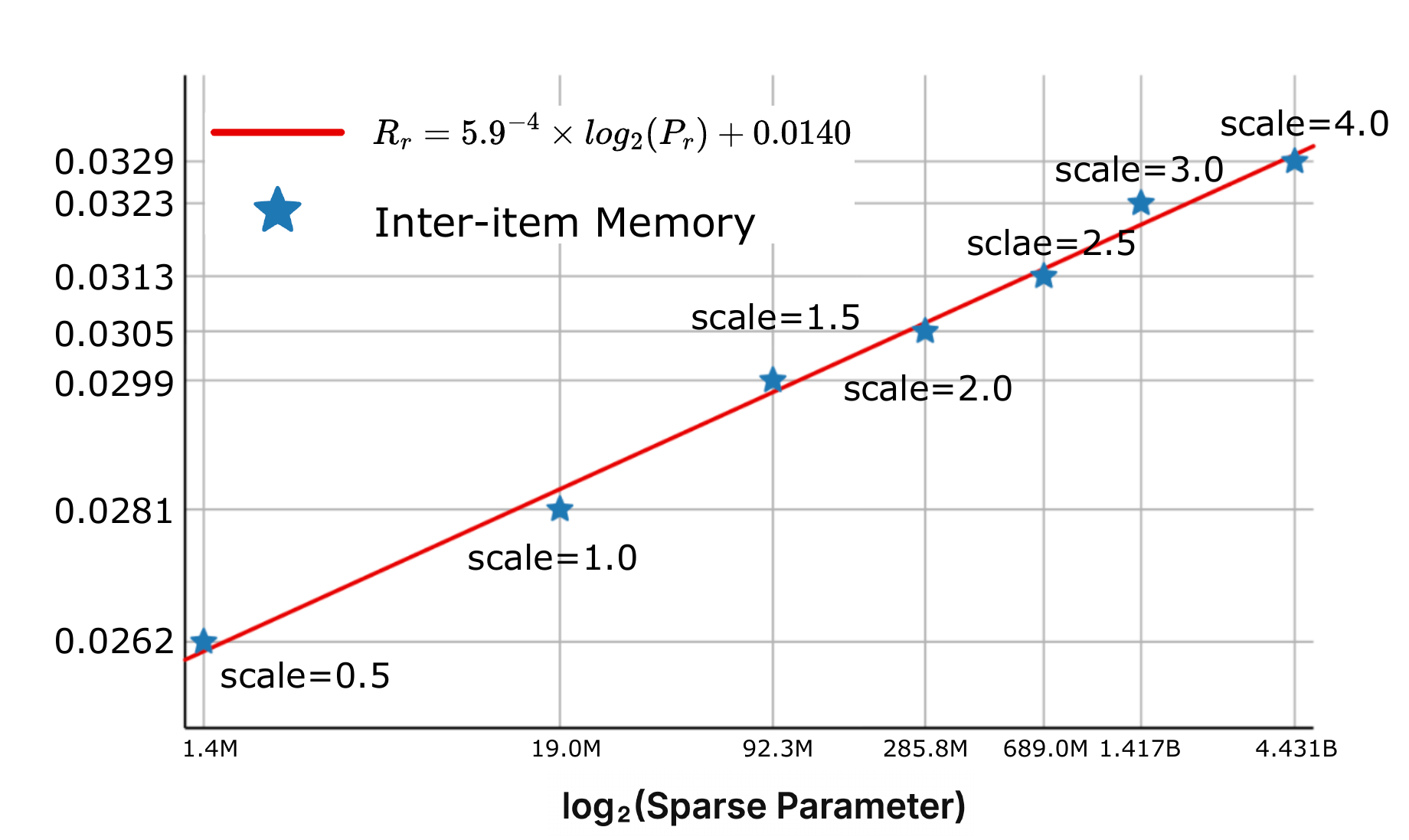}
        \caption{Inter-item Memory}
        \label{fig:inter_scaling}
    \end{subfigure}

    \caption{Scaling analysis of dual-level conditional memory on the Yelp dataset.}
    \label{fig:engram_scaling}
     \vspace{-3mm}
\end{figure}
\subsection{Scaling Analysis (RQ3)}
\label{sec:scaling}

We also examine whether expanding the two memories, \ie intra-item and inter-item memories, could compile to a similar scaling law. The results are plotted in Figure~\ref{fig:engram_scaling}. For both memories, we observe a consistent scaling phenomenon for the two conditional memories, consistent with previous findings~\cite{cheng2026conditional}. These results confirm the scalability of the proposed dual-level conditional memory: \textbf{increasing sparse capacity reduces hash collisions and continuously improves the quality of capturing the structured SID relations, for both the representation construction and LLM decoding.} More details are provided in Appendix~\ref{sec:details_scala}.

\begin{figure*}[!t]
	\centering
	\includegraphics[width = \linewidth]{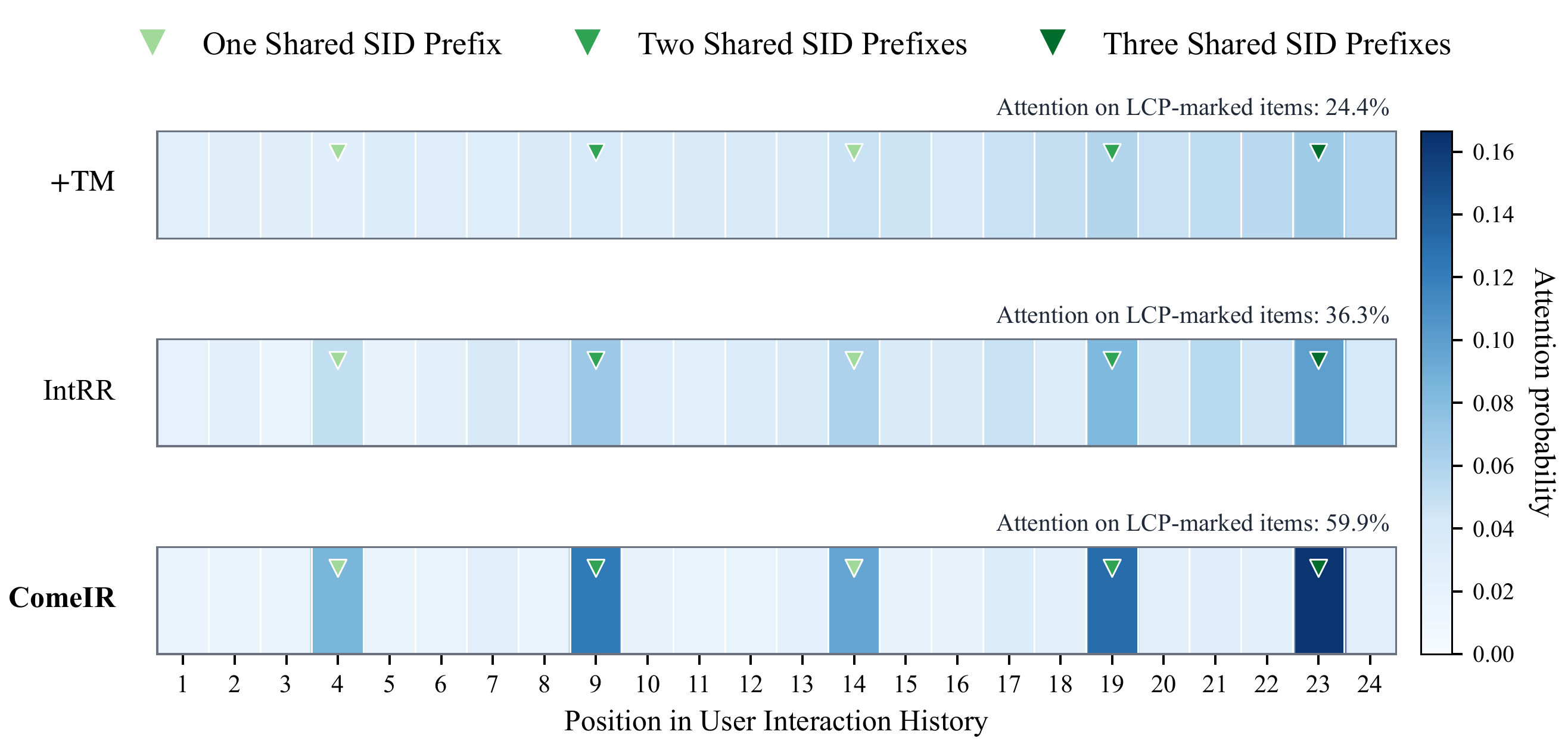}

	\caption{Attention distributions over historical item representations across different representation constructors (all under Vanilla GR). Light, medium, and dark green markers denote longest-common-prefix (LCP) lengths of one, two, and three between a historical item's SID and the ground-truth target SID, respectively.}\label{fig:case}
    \vspace{-4mm}
\end{figure*}
\subsection{Efficiency Analysis (RQ4)}
As previously discussed, the item-level representation significantly reduces the length of the GR input from $N \times L$ to $N$. To further estimate this, we report the inference latency of \name~in Table~\ref{Tab: LLM_Inference_LT}. The results show that, by optimizing the representation stage rather than quantization or generation, our proposed \name~ achieves a remarkable $\textbf{2.5} \times$ average speedup through length reduction compared to the flattened setting, even with the efficient NEZHA architecture. While lagging a little behind the naive token merge, it still achieves better latency compared to the powerful baseline IntRR. \textbf{This phenomenon further validates that employing \name~in the current GR pipeline can achieve better balance between effectiveness and efficiency.} More details can be found in Appendix~\ref{sec:efficiency_app}.

\subsection{Case Study (RQ5)}
To further validate the effectiveness of \name, we conduct a case study on a randomly selected user 14310, who owns 24 historical interactions from Amazon Industrial. We compare \textit{+ TM}, IntRR, and \name~under the same Vanilla GR backbone, and visualize the attention from the final user state to each historical item representation. The scores are averaged across all heads in the last four Transformer layers. The LCP markers are added only for post-hoc analysis and are not provided to the model.

As shown in Figure~\ref{fig:case}, \textbf{\name~focuses more strongly on target-relevant historical items.} The total attention assigned to LCP-marked items increases from 24.4\% for \textit{+ TM} and 36.3\% for IntRR to 59.9\% for \name. \textbf{Moreover, \name~preserves the hierarchical SID relations.} It assigns the highest attention to the historical item with a three-level SID match, while also emphasizing items with one- and two-level prefix matches. These observations suggest that \name~retains structured SID evidence in item-level representations, enabling the LLM to aggregate relevant historical interests more selectively. For a detailed case at the representation stage, please refer to Appendix~\ref{sec:representation_case_app}.

\section{Related Works}
\subsection{Generative Recommendation.} Generative recommendation (GR) reformulates item retrieval as autoregressive identifier generation conditioned on user histories. P5~\cite{geng2022recommendation} casts recommendation tasks into language processing, and TIGER~\cite{rajput2023recommender} establishes the representative Semantic ID (SID) pipeline, where each item is encoded as SID, and the LLM directly generates the next item. Following the quantization-generation pipeline~\cite{zhai2024actions,yang2024unifying,deng2025onerec,zhou2025onerec,lin2024efficient,mekonnen2026parametric}, a line of research improve SID construction with content, collaborative signals, or task-aware tokenization~\cite{hou2023learning,wang2024eager,wang2024learnable,zhu2024cost,liu2024end}, while others study SID-language alignment, long-SID generation, inductive decoding, and SID redistribution~\cite{zheng2024adapting,hou2025generating,ding2026inductive,wang2026intrr}. 

While these works advance SID and generator design, how to derive better representations that are semantics-recoverable, relation-aware, and compatible with token-level prediction for GR input remains underexplored. To address this, we thus propose Item-guided Token Scoring to estimate the contribution of each code, supervised by the initial item semantic embeddings, for recovering the lost semantics. And leveraging a dual-level conditional memory to maintain the structured SID relations for different patterns. 

\subsection{Conditional Memory.} Memory-based modeling reuses recurring patterns through explicit or implicit storage. Classical $n$-gram language models store local statistics~\cite{katz1987estimation,kneser1995improved}, general $n$-gram and hash embeddings encode reusable evidence as compact vectors~\cite{zhao2017ngram2vec,tito2017hash}, and modern language models exhibit memory-like behavior through key-value feed-forward layers or retrieval-augmented generation~\cite{geva2021transformer,cheng2023lift}. Conditional computation and Mixture-of-Experts improve scaling by sparse activation, while Engram~\cite{cheng2026conditional} introduces conditional memory as sparse lookup over $n$-gram patterns, with follow-up work studying its serving and indexing properties~\cite{lin2026collision}. 

While existing work uses conditional memory to capture linguistic patterns, its role in preserving fine-grained SID evidence, such as inter-item and intra-item token relations, regardless of the representation stage or the generation stage, remains underexplored. We thus propose our dual-level conditional memory with Memory Merge and a Memory-restoring Prediction Head to optimize representation and generation, respectively.
\section{Conclusion}
In this paper, we identify representation construction as a key bottleneck in current GR, where existing representation constructors face two severe challenges, \ie Loss of item semantics and insufficient modeling of structured SID relations. Consequently, we propose \name, which contains several innovations: 1) Item-guided token scoring, which estimates the contribution of each code, eliminating the loss of item semantics. 2) Dual-level conditional memories, which specially design two separate n-gram hash memories, providing a foundation for preserving  SID structured relations. 3) The Memory Merge, which combines the proposed memories with the scored tokens, deriving the fine-grained representation. 4) The Memory-Restoring Prediction Head further leverage these memories to enhance subsequent generation for each SID prefix during autoregressive decoding. Extensive experiments demonstrate the effectiveness, flexibility, and scalability of the proposed \name.
\bibliographystyle{ACM-Reference-Format}
\bibliography{8Reference}
\clearpage
\appendix
\section{Supplement to Method}\label{sec:method_supp}
This section supplements Section~\ref{sec:method} with the addressing, table construction, and architecture-specific details omitted from the main text. We retain the notation introduced in the general conditional-memory formulation.

\subsection{General Conditional Memory}\label{sec:engram_general}
Equation~\eqref{equ:engram_read} defines the shared conditional-memory read. Here, we specify the multi-head hash mapping and the low-rank address adaptation. For a discrete sequence $\boldsymbol x=(x_1,\ldots,x_T)$, the order-$o$ hash at position $n$ is
\begin{equation}\label{equ:app_engram_hash}
\phi_{\ell,o,k}(x_{n-o+1:n})
=
\left(
\bigoplus_{j=0}^{o-1}x_{n-j}m_{j}^{\ell}
\right)\bmod H_{\ell,o,k},
\end{equation}
where unavailable left positions are padded, $m_j^{\ell}$ is a fixed odd multiplier, and the hash heads use distinct prime moduli $H_{\ell,o,k}$. Different SID levels use independently seeded hash mappings. For each order, the head-wise lookup is
\begin{equation}\label{equ:app_engram_address}
\boldsymbol a_{n,\ell,o}^{X}
=
\mathop{\Vert}_{k=1}^{K_X}
\boldsymbol M_{\ell,o,k}^{X}
[\phi_{\ell,o,k}^{X}(\boldsymbol x)],
\end{equation}
and Equation~\eqref{equ:general_memory_address} concatenates these addresses over $o\in\mathcal O_\ell^X$.

For Intra-item Memory, each addressed vector additionally contains a low-rank residual:
\begin{equation}\label{equ:app_intra_lora}
\widehat{\boldsymbol M}_{\ell,o,k}^{S}[h]
=
\boldsymbol M_{\ell,o,k}^{S}[h]
+\boldsymbol B_{\ell}^{S}
\boldsymbol A_{\ell,o,k}^{S}[h],
\end{equation}
where $\boldsymbol A[h]\in\mathbb R^{r_S}$ and $\boldsymbol B\in\mathbb R^{d_h\times r_S}$. The low-rank branch is initialized near zero and trained jointly with the sparse tables. Inter-item Memory directly uses the retrieved table vectors before the projections in Equation~\eqref{equ:engram_read}.

\subsection{Intra-item Memory}\label{sec:intra_engram}
For Equations~\eqref{equ:intra_prefix_encoding}--\eqref{equ:intra_memory}, we create one prefix-table group for each level $t=2,\ldots,L$. The encoded prefix $\kappa_{n,t}$ is treated as one discrete address unit and retrieves the low-rank-adapted vectors in Equation~\eqref{equ:app_intra_lora}. The level-$\ell$ composition adapter has input dimension $(\ell-1)d$ and combines $\boldsymbol z_{n,2}^S,\ldots,\boldsymbol z_{n,\ell}^S$ before RMS normalization.

During SID prediction, the generated prefixes directly query the same table groups. At level $\ell$, prefixes of lengths $2,\ldots,\ell-1$ are available; multiple retrieved signals are averaged before the residual gate in Equation~\eqref{equ:decode_intra_refine}. For $L=3$, this prefix-memory refinement is activated when predicting the third code.

\subsection{Inter-item Memory}\label{sec:inter_engram}
For Equations~\eqref{equ:inter_prefix_sequence}--\eqref{equ:inter_level_aggregation}, the maximum memory orders are $(O_1^T,O_2^T,O_3^T)=(3,2,1)$. At level $\ell$, an order-$o$ key contains the most recent $o$ elements of the rearranged sequence $\mathcal C^{\leq\ell}$ ending at the current position. The local query window in Equation~\eqref{equ:inter_query} is set to $w=5$.

After the conditional read, $\operatorname{ShortConv}_\ell$ applies a depth-wise convolution with kernel size four and dilation $O_\ell^T$. The level-attention network in Equation~\eqref{equ:inter_level_aggregation} is shared across SID levels and maps each $\boldsymbol\eta_{n,\ell}^{T}$ to a scalar before softmax normalization. The full tensor $\{\boldsymbol\eta_{n,\ell}^{T}\}_{n,\ell}$ is retained for the historical aggregation in Equation~\eqref{equ:decode_inter_context}.

\subsection{Memory-restoring Prediction Head}\label{sec:prediction_head_supp}
The prediction head in Section~\ref{sec:prediction_head} uses the final historical hidden state $\boldsymbol h_u$ to restore prefix-composition and historical-transition evidence. For the NEZHA architecture, we additionally append $L$ trainable level prompts:
\begin{equation}\label{equ:app_prompt_input}
[\boldsymbol H^I;\boldsymbol H^{\mathrm{ph}}]
=
\operatorname{LLM}([
\boldsymbol R^I;
\boldsymbol e_1^{\mathrm{ph}};\ldots;
\boldsymbol e_L^{\mathrm{ph}}]),
\end{equation}
where $\boldsymbol h_u$ is the last valid state in $\boldsymbol H^I$ and $\boldsymbol h_\ell^{\mathrm{ph}}$ is the $\ell$-th state in $\boldsymbol H^{\mathrm{ph}}$. Let $\boldsymbol h_u^1=\boldsymbol h_u$. NEZHA replaces the Vanilla state $\boldsymbol h_u$ in the generated-prefix refinement and prediction fusion with
\begin{equation}\label{equ:app_nezha_state}
\boldsymbol\zeta_\ell
=\boldsymbol h_u^\ell+\boldsymbol h_\ell^{\mathrm{ph}}.
\end{equation}
Thus, Equation~\eqref{equ:decode_intra_refine} is evaluated using $\boldsymbol\zeta_\ell$ instead of $\boldsymbol h_u$, producing $\widehat{\boldsymbol\zeta}_\ell$. Equation~\eqref{equ:decode_fusion} then replaces $\widehat{\boldsymbol h}_{u,\ell}$ with $\widehat{\boldsymbol\zeta}_\ell$. The historical inter-memory context in Equation~\eqref{equ:decode_inter_context} is pooled using the original final-history state $\boldsymbol h_u$.

After selecting code $c_{N+1}^\ell$, NEZHA combines its embedding with the next-level inter context and updates the path-specific state:
\begin{equation}\label{equ:app_nezha_transition}
\begin{aligned}
\boldsymbol g_\ell^{\mathrm{tr}}
&=\sigma\!\left(
\boldsymbol W_{g,\ell}^{\mathrm{tr}}
[\boldsymbol e_{c_{N+1}^\ell};
\bar{\boldsymbol\eta}_{u,\ell+1}^{T}]
+\boldsymbol b_{g,\ell}^{\mathrm{tr}}
\right),\\
\boldsymbol d_\ell^{\mathrm{tr}}
&=\boldsymbol g_\ell^{\mathrm{tr}}\odot
\boldsymbol W_{\Delta,\ell}
[\boldsymbol e_{c_{N+1}^\ell};
\bar{\boldsymbol\eta}_{u,\ell+1}^{T}]
+(1-\boldsymbol g_\ell^{\mathrm{tr}})\odot
\boldsymbol e_{c_{N+1}^\ell},\\
\boldsymbol h_u^{\ell+1}
&=\operatorname{GRU}_\ell(
\boldsymbol d_\ell^{\mathrm{tr}},\boldsymbol h_u^\ell),
\qquad \ell<L.
\end{aligned}
\end{equation}
The recurrent transition propagates the selected-code and transition evidence across SID levels. Teacher forcing supplies the selected code during training, while inference maintains one recurrent state for each NEZHA beam path.

\subsection{Training and Inference}\label{sec:train_infer}
We summarize the training and inference procedures in Algorithm~\ref{alg:training_inference}. The representation and memory operations are shared across architectures, while the decoding state follows the corresponding prediction backbone.
\begin{algorithm}[t]
\caption{Training and inference of \name}
\label{alg:training_inference}
\begin{algorithmic}[1]
\Require History $\mathcal S_u$, architecture $A\in\{\mathrm{Vanilla},\mathrm{NEZHA}\}$, target SID $\boldsymbol c_{N+1}$ in training, catalog trie $\mathcal T$, mode $z$, beam size $B$
\State Construct $\boldsymbol s_n^0$ for every history item by Eq.~\eqref{equ:sid_token_scoring}
\State Retrieve $\boldsymbol\eta_{n,\ell}^{S}$ and $\boldsymbol\eta_{n,\ell}^{T}$ by Eqs.~\eqref{equ:intra_memory} and~\eqref{equ:inter_memory}
\State Construct $\boldsymbol r_n^I$ by Eq.~\eqref{equ:token_merge}; cache $\boldsymbol\eta_{n,\ell}^{T}$
\If{$A=\mathrm{Vanilla}$}
    \State Run $\boldsymbol H^I=\operatorname{LLM}(\boldsymbol R^I)$ and set $\boldsymbol h_u=\boldsymbol H_N^I$
\Else
    \State Append the level prompts and obtain $\boldsymbol h_u,\boldsymbol h_1^{\mathrm{ph}},\ldots,\boldsymbol h_L^{\mathrm{ph}}$ by Eq.~\eqref{equ:app_prompt_input}
    \State Initialize $\boldsymbol h_u^1=\boldsymbol h_u$
\EndIf
\State Pool $\bar{\boldsymbol\eta}_{u,\ell}^{T}$ for every level using $\boldsymbol h_u$ by Eq.~\eqref{equ:decode_inter_context}
\State Initialize one empty path with score zero
\For{$\ell=1$ to $L$}
    \For{each active path}
        \If{$A=\mathrm{Vanilla}$}
            \State Set the prediction anchor to $\boldsymbol h_u$
        \Else
            \State Set the prediction anchor to $\boldsymbol\zeta_\ell=\boldsymbol h_u^\ell+\boldsymbol h_\ell^{\mathrm{ph}}$
        \EndIf
        \State Refine the anchor from the already generated prefix by Eq.~\eqref{equ:decode_intra_refine}
        \State Fuse $\bar{\boldsymbol\eta}_{u,\ell}^{T}$ and compute level logits by Eq.~\eqref{equ:decode_fusion}
        \If{$z=\mathrm{Train}$}
            \State Accumulate the ground-truth code loss and select it by teacher forcing
        \Else
            \State Expand with top-scoring codes and mask extensions invalid under $\mathcal T$
        \EndIf
        \If{$A=\mathrm{NEZHA}$ and $\ell<L$}
            \State Update the path state using Eq.~\eqref{equ:app_nezha_transition}
        \EndIf
    \EndFor
    \If{$z=\mathrm{Infer}$}
        \State Retain the top-$B$ paths by accumulated log-probability
    \EndIf
\EndFor
\State Return the cross-entropy loss or the ranked catalog-valid SIDs
\end{algorithmic}
\end{algorithm}

\section{Experimental Settings}\label{sec:experimental_settings}
\subsection{Dataset Statistics}\label{sec:data}
We follow the preprocessing of prior GR studies~\cite{hou2023learning,rajput2023recommender}; statistics are reported in Table~\ref{tab:dataset}.
\begin{table}[t]
\centering
\caption{Dataset statistics after preprocessing.}\label{tab:dataset}
\small
\setlength{\tabcolsep}{3.5pt}
\begin{tabular}{ccccc}
\toprule
Dataset & \#Users & \#Items & Sparsity & Avg. len.\\
\midrule
Yelp & 15,720 & 11,383 & 99.89\% & 12.23\\
Industrial & 27,190 & 28,461 & 99.92\% & 6.56\\
Instrument & 40,644 & 30,676 & 99.97\% & 8.01\\
\bottomrule
\end{tabular}
\end{table}

\subsection{Detailed Implementation}\label{sec:implementation_details}
Experiments run on an AMD EPYC 9745 platform with two NVIDIA RTX PRO 6000 GPUs (Blackwell, 96 GB), Python 3.11, and PyTorch 2.10. The cached $\boldsymbol m_n$ is the embedding supplied to the quantizer when constructing the SID. A learned spectral reweighting module can be applied when a precomputed SVD file is available.

\subsubsection{Representation Construction}\label{sec:representation_impl}
We use $L=3$ SID levels and codebook size $C=128$ at each level. The maximum capacity of one hash table is $H_{\max}=20{,}000{,}000$, and the encoded level-$\ell$ prefix domain is $D_\ell=\min(C^\ell,2{,}097{,}152)$. The concatenated address dimension per order is 256: Intra-item Memory uses $K_S=2$ heads of dimension 128, while Inter-item Memory uses $K_T=4$ heads of dimension 64. We set the intra-item low-rank adapter to $r_S=32$, corresponding to a rank ratio of $0.25$.

\noindent\textbf{Table allocation.}
For $\mathcal O_t^S=\{1,\ldots,\min(t,3)\}$ and $B^S(s_S)=128s_S$,
\begin{equation}\label{equ:app_intra_bucket}
\Gamma_{t,o}^S(s_S)=\max\!\left\{2,\min\!\left(
\lfloor(B^S(s_S))^o\rfloor,C^o,H_{\max}\right)\right\},
\end{equation}
\begin{equation}\label{equ:app_intra_sparse_params}
P_S(s_S)=\frac{256}{K_S}\sum_{t=2}^{L}
\sum_{o\in\mathcal O_t^S}\sum_{k=1}^{K_S}H_{t,o,k}^S.
\end{equation}
For $(O_1^T,O_2^T,O_3^T)=(3,2,1)$, $\mathcal O_\ell^T=\{1,\ldots,O_\ell^T\}$, and $B_\ell^T(s_T)=(16s_T)^\ell$,
\begin{equation}\label{equ:app_inter_bucket}
\Gamma_{\ell,o}^T(s_T)=\max\!\left\{2,\min\!\left(
\lfloor(B_\ell^T(s_T))^o\rfloor,D_\ell^o,H_{\max}\right)\right\},
\end{equation}
\begin{equation}\label{equ:app_inter_sparse_params}
P_T(s_T)=\frac{256}{K_T}\sum_{\ell=1}^{L}
\sum_{o\in\mathcal O_\ell^T}\sum_{k=1}^{K_T}H_{\ell,o,k}^T.
\end{equation}
Each head is assigned a distinct prime capacity at or below the corresponding target whenever possible. The sparse counts above exclude dense projections, low-rank adapters, and normalization parameters. The default scales are $s_S=1.0$ and $s_T=2.0$. Inter-item retrieval uses a local query window of five items and a depth-wise convolution with kernel size four and dilation $O_\ell^T$.

\subsubsection{Generation}\label{sec:generation_impl}
For the Vanilla formulation in the main text, the LLM receives the $N$ item representations, and its final valid hidden state is reused as the prediction anchor at every SID level. The NEZHA variant additionally appends three trainable level prompts, extracts their hidden states, and maintains the recurrent transition in Equation~\eqref{equ:app_nezha_transition}. Both variants use one linear classifier per SID level and apply the catalog trie after computing level logits. NEZHA additionally maintains a recurrent state for every surviving beam path.

\noindent\textbf{Optimization settings.}
We train with bfloat16, a base learning rate of $10^{-5}$, weight decay $10^{-4}$, batch size 64 per device, two gradient-accumulation steps, and 10K optimization steps. Evaluation and checkpointing are performed every 100 steps. Sparse embeddings, context gates, adapters, merge modules, recurrent transitions, and prediction-fusion modules use dedicated learning-rate groups; the scale-dependent sparse-memory groups are further multiplied by $\sqrt{\max(1,s_T)}$ or $\sqrt{\max(1,s_S)}$. We run the same configuration with the reported random seeds.

\section{Extra Experimental Results}
\subsection{Ablation Study}\label{sec:ablation_supp}
Tables~\ref{Tab: Ablation_Industrial}--\ref{Tab: Ablation_Instrument} extend Table~\ref{Tab: Ablation_Main} under Qwen3-0.6B and RQ-VAE. \textbf{R}-intra/inter remove the corresponding representation-side branch. \textbf{G}-intra removes generated-prefix refinement, while \textbf{G}-inter removes historical inter-memory fusion. \textbf{Finding:} Both representation construction and prediction-state refinement are beneficial, and the matched dense head remains inferior.

\begin{table}[t]
\centering
\caption{Ablation results on Amazon Industrial.}
\label{Tab: Ablation_Industrial}
\small
\setlength{\tabcolsep}{3.2pt}
\renewcommand{\arraystretch}{0.78}
\resizebox{\columnwidth}{!}{
\begin{tabular}{ccccc}
\toprule
\textbf{Variant} & \textbf{H@5} & \textbf{H@10} & \textbf{N@5} & \textbf{N@10}\\
\midrule
\cellcolor{cyan!20}\textbf{\name} & \cellcolor{cyan!20}\textbf{0.1031} & \cellcolor{cyan!20}\textbf{0.1321} & \cellcolor{cyan!20}\textbf{0.0792} & \cellcolor{cyan!20}\textbf{0.0881}\\
\textit{+ TM} & 0.0976 & 0.1271 & 0.0742 & 0.0841\\
\midrule
\textit{w/o} Token Scoring & 0.1001 & 0.1292 & 0.0767 & 0.0852\\
\textit{w/o} Memory Merge & 0.1008 & 0.1297 & 0.0774 & 0.0860\\
\textit{w/o} \textbf{R}-intra & 0.1012 & 0.1300 & 0.0778 & 0.0864\\
\textit{w/o} \textbf{R}-inter & 0.1020 & 0.1310 & 0.0785 & 0.0872\\
\textit{w/o} \textbf{R}-Memory & 0.0998 & 0.1287 & 0.0763 & 0.0854\\
\textit{w/o} \textbf{G}-intra & 0.1005 & 0.1293 & 0.0770 & 0.0855\\
\textit{w/o} \textbf{G}-inter & 0.1016 & 0.1306 & 0.0781 & 0.0868\\
\textit{w/o} \textbf{G}-Memory & 0.0992 & 0.1282 & 0.0758 & 0.0850\\
\textit{w/o} All Memory & 0.0988 & 0.1280 & 0.0754 & 0.0847\\
Matched Dense Head & 0.1003 & 0.1295 & 0.0769 & 0.0857\\
\bottomrule
\end{tabular}}
\vspace{-2mm}
\end{table}

\begin{table}[t]
\centering
\caption{Ablation results on Amazon Instrument.}
\label{Tab: Ablation_Instrument}
\small
\setlength{\tabcolsep}{3.2pt}
\renewcommand{\arraystretch}{0.78}
\resizebox{\columnwidth}{!}{
\begin{tabular}{ccccc}
\toprule
\textbf{Variant} & \textbf{H@5} & \textbf{H@10} & \textbf{N@5} & \textbf{N@10}\\
\midrule
\cellcolor{cyan!20}\textbf{\name} & \cellcolor{cyan!20}\textbf{0.0834} & \cellcolor{cyan!20}\textbf{0.1021} & \cellcolor{cyan!20}\textbf{0.0709} & \cellcolor{cyan!20}\textbf{0.0772}\\
\textit{+ TM} & 0.0788 & 0.0966 & 0.0672 & 0.0719\\
\midrule
\textit{w/o} Token Scoring & 0.0808 & 0.0991 & 0.0685 & 0.0744\\
\textit{w/o} Memory Merge & 0.0815 & 0.0996 & 0.0693 & 0.0752\\
\textit{w/o} \textbf{R}-intra & 0.0817 & 0.1000 & 0.0695 & 0.0756\\
\textit{w/o} \textbf{R}-inter & 0.0825 & 0.1011 & 0.0702 & 0.0765\\
\textit{w/o} \textbf{R}-Memory & 0.0804 & 0.0985 & 0.0682 & 0.0739\\
\textit{w/o} \textbf{G}-intra & 0.0811 & 0.0994 & 0.0689 & 0.0749\\
\textit{w/o} \textbf{G}-inter & 0.0821 & 0.1007 & 0.0699 & 0.0761\\
\textit{w/o} \textbf{G}-Memory & 0.0799 & 0.0980 & 0.0679 & 0.0735\\
\textit{w/o} All Memory & 0.0795 & 0.0976 & 0.0676 & 0.0730\\
Matched Dense Head & 0.0810 & 0.0993 & 0.0688 & 0.0748\\
\bottomrule
\end{tabular}}
\vspace{-2mm}
\end{table}

\subsection{Representation Case Study}\label{sec:representation_case_app}
\begin{figure*}[!t]
	\centering
	\includegraphics[width =\linewidth]{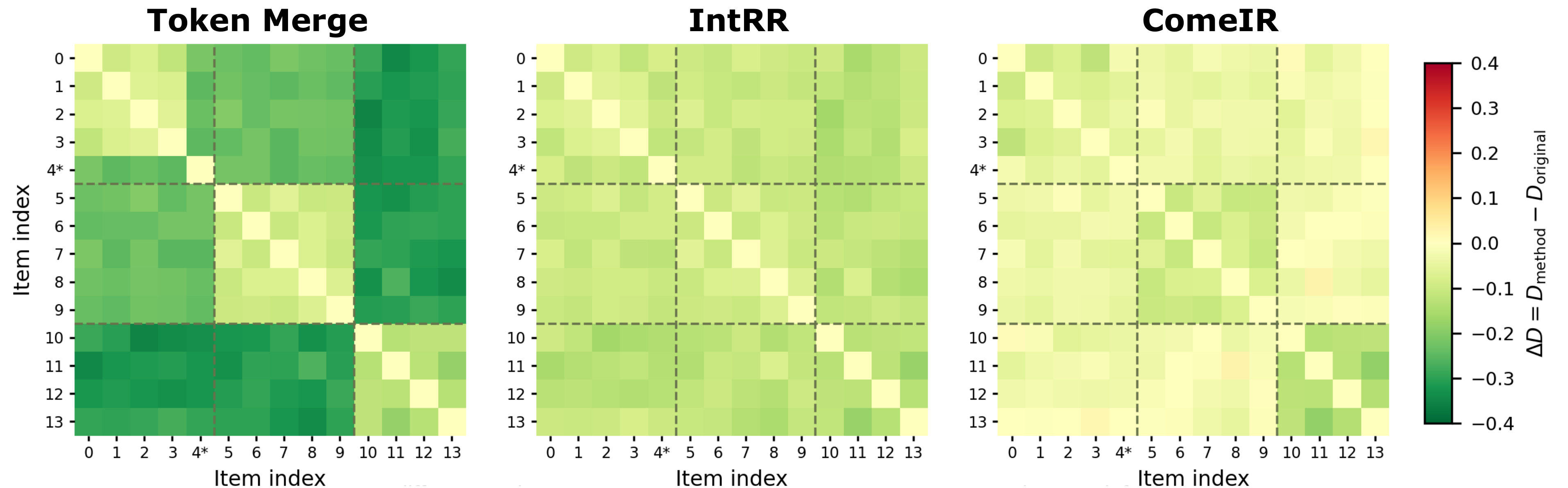}

	\caption{Representation-distance changes for 14 Amazon Industrial items. Indices $[0,3]$, $[5,9]$, and $[10,13]$ form three categories; item 4 belongs to another category.}\label{fig:representation_case_app}
	\Description{Three heatmaps compare pairwise representation-distance changes for Token Merge, IntRR, and ComeIR.}
    \vspace{-5mm}
\end{figure*}
Figure~\ref{fig:representation_case_app} reports $\Delta D=D_{\mathrm{method}}-D_{\mathrm{original}}$, where $D=(1-\cos(\boldsymbol a,\boldsymbol b))/2$. Token Merge contracts cross-category distances and IntRR partly alleviates this effect. 

\noindent\textbf{Finding: ComeIR better preserves category separation}, including the boundary between item 4 and items 0--3.

\subsection{Detailed Efficiency Analysis}\label{sec:efficiency_app}
\begin{table}[t]
\centering
\caption{Efficiency with Qwen3-0.6B and Vanilla GR (batch size 32; beam size 20).}
\label{tab:detailed_efficiency}
\small
\setlength{\tabcolsep}{3.4pt}
\renewcommand{\arraystretch}{0.78}
\resizebox{\linewidth}{!}{
\begin{tabular}{cc|c|cc}
\toprule
\multirow{2}{*}{Dataset} & \multirow{2}{*}{Method} & Train & \multicolumn{2}{c}{Peak memory (GB)}\\
\cmidrule(lr){4-5}
& & samples/s $\uparrow$ & Train $\downarrow$ & Infer. $\downarrow$\\
\midrule
\multirow{3}{*}{Yelp} & Flattened & 552 & 10.97 & 8.44\\
& \textit{+ TM} & 1185 & 9.26 & 3.77\\
& \cellcolor{cyan!20}\textbf{\name} & \cellcolor{cyan!20}1046 & \cellcolor{cyan!20}19.48 & \cellcolor{cyan!20}8.89\\
\midrule
\multirow{3}{*}{Industrial} & Flattened & 617 & 10.18 & 5.89\\
& \textit{+ TM} & 1308 & 9.13 & 3.66\\
& \cellcolor{cyan!20}\textbf{\name} & \cellcolor{cyan!20}1139 & \cellcolor{cyan!20}18.41 & \cellcolor{cyan!20}6.51\\
\midrule
\multirow{3}{*}{Instrument} & Flattened & 499 & 12.10 & 7.26\\
& \textit{+ TM} & 1224 & 10.03 & 3.92\\
& \cellcolor{cyan!20}\textbf{\name} & \cellcolor{cyan!20}1077 & \cellcolor{cyan!20}19.88 & \cellcolor{cyan!20}7.85\\
\bottomrule
\end{tabular}}
\vspace{-2mm}
\end{table}
\noindent\textbf{Finding 1:} Reducing the input from $3N$ to $N$ yields $1.85$--$2.16\times$ the throughput of Flattened; \textit{+ TM} remains faster without sparse retrieval. 

\noindent\textbf{Finding 2:} Retrieved rows and optimizer states raise peak memory, so \name{} retains the latency benefit at a higher memory cost.
\subsection{Details of Scalability Analysis}\label{sec:details_scala}
We scale one memory at a time, fixing the other at its default. Figure~\ref{fig:engram_scaling} reports $P_S(s_S)$ and $P_T(s_T)$ from Equations~\eqref{equ:app_intra_sparse_params}--\eqref{equ:app_inter_sparse_params}.

For intra-item memory, $s_S\in\{0.125,0.25,0.5,0.75,1.0\}$ gives $B^S\in\{16,32,64,96,128\}$; counts are capped by $C^o$ and $H_{\max}$ and expanded into two heads of dimension 128. For inter-item memory, $s_T\in\{0.5,1.0,\ldots,4.0\}$ and $B_\ell^T=(16s_T)^\ell$; counts are capped by $D_\ell^o$ and $H_{\max}$ and expanded into four heads of dimension 64.

\noindent\textbf{Finding 1: Intra-item memory shows diminishing returns.} H@5 improves with $s_S$ but begins to saturate near $0.75$ as most effective collisions in the bounded composition space are resolved.

\noindent\textbf{Finding 2: Inter-item memory benefits from larger capacities.} Its broader transition space grows from approximately $1.4\,\mathrm{M}$ to $4.431\,\mathrm{B}$ sparse parameters and continues improving within the tested range.

\end{document}